\documentclass[10pt,letterpaper]{article}
\usepackage[top=0.85in,left=2.75in,footskip=0.75in]{geometry}

\usepackage{amsmath,amssymb}

\usepackage[hidelinks]{hyperref} 
\usepackage{colortbl}
\usepackage{float}
\usepackage{supertabular}
\usepackage{rotating}
\usepackage{booktabs}
\usepackage{ragged2e}
\usepackage[longtable]{multirow}
\usepackage{longtable}
\usepackage{array}
\usepackage{graphicx}
\usepackage{verbatim} 
\usepackage{interval}
\usepackage{amsmath,amssymb,amsfonts}
\usepackage{algorithmic}
\usepackage{subfigure}
\usepackage{adjustbox}
\usepackage{caption}
\usepackage{textcomp}
\usepackage{multicol}
\usepackage{setspace}
\restylefloat{figure}
\usepackage{textcomp}
\usepackage{threeparttable}
\usepackage{multicol}
\usepackage{interval}
\usepackage{float}
\usepackage{tabularx}
\usepackage{tikz}
\usepackage{caption}
\usepackage{wrapfig}
\usepackage{wrapfig, subcaption, setspace, booktabs}
\usepackage{multirow}
\usetikzlibrary{shapes, arrows, positioning}
\usepackage{scalerel}

\usepackage{changepage}

\usepackage{textcomp,marvosym}

\usepackage{cite}

\usepackage{nameref,hyperref}

\usepackage[right]{lineno}

\usepackage[nopatch=eqnum]{microtype}
\DisableLigatures[f]{encoding = *, family = * }

\usepackage{array}

\newcolumntype{+}{!{\vrule width 2pt}}

\newlength\savedwidth

\raggedright
\usepackage[aboveskip=1pt,labelfont=bf,labelsep=period,justification=raggedright,singlelinecheck=off]{caption}

\makeatletter
\renewcommand{\@biblabel}[1]{\quad#1.}
\makeatother

\usepackage{lastpage,fancyhdr,graphicx}
\usepackage{epstopdf}
\fancyheadoffset[L]{2.25in}
\fancyfootoffset[L]{2.25in}
\begin{document}
\vspace*{0.2in}

\begin{flushleft}
{\Large
\textbf\newline{Analysis of Internet of Things Implementation Barriers in the Cold Supply Chain: An Integrated ISM--MICMAC and DEMATEL Approach} 
}
\newline
\\
Kazrin Ahmad\textsuperscript{1\ddag},
Md. Saiful Islam\textsuperscript{1},
Md Abrar Jahin\textsuperscript{1\ddag},
and M. F. Mridha\textsuperscript{2*}

\bigskip
\textbf{1} Department of Industrial Engineering and Management, Khulna University of Engineering and Technology (KUET), Khulna 9203, Bangladesh
\\
\textbf{2} Department of Computer Science, American International University-Bangladesh, Dhaka 1229, Bangladesh
\\

\bigskip


%
%
\ddag Co-first author(s): Kazrin Ahmad and Md Abrar Jahin. These authors contributed equally to this work.

%
%






*firoz.mridha@aiub.edu

\end{flushleft}


\begin{justify}
\section*{Abstract}
Integrating Internet of Things (IoT) technology inside the cold supply chain can enhance transparency, efficiency, and quality, optimize operating procedures, and increase productivity. The integration of the IoT in this complicated setting is hindered by specific barriers that require thorough examination. Prominent barriers to IoT implementation in a cold supply chain, which is the main objective, are identified using a two--stage model. After reviewing the available literature on IoT implementation, 13 barriers were identified. The survey data were cross--validated for quality, and Cronbach's alpha test was employed to ensure validity. This study applies the interpretative structural modeling technique in the first phase to identify the main barriers. Among these barriers, ``regulatory compliance" and ``cold chain networks" are the key drivers of IoT adoption strategies. MICMAC's driving and dependence power element categorization helps evaluate barrier interactions. In the second phase of this study, a decision-making trial and evaluation laboratory methodology was employed to identify causal relationships between barriers and evaluate them according to their relative importance. Each cause is a potential drive, and if its efficiency can be enhanced, the system benefits as a whole. The findings provide industry stakeholders, governments, and organizations with significant drivers of IoT adoption to overcome these barriers and optimize the utilization of IoT technology to improve the effectiveness and reliability of the cold supply chain.



\section{Introduction}
\label{sec:introduction}
Recent developments in Internet of Things (IoT) technology have the potential to significantly change several industries via the introduction of data-driven insights, increased connectivity, and a replacement of long-established ways of doing things.
The supply chain (SC) is a network of related entities that includes every procedure step, from the procurement of raw materials or components to the consummation of the finished good. Supply chains in the modern day have developed into intricate value networks and are now an important differentiator for businesses. A growing number of barriers make it harder than ever to track items and goods as they go through the value chain and confirm where they came from \cite{rejeb2019leveraging}. Cold SC (CSC) is a procedure that includes activities regulated by temperature and a specialized logistical infrastructure designed to maintain product quality and temperature consistency throughout its entire transit, beginning with the point of origin in manufacturing or procurement and ending with its eventual delivery to the final consumer \cite{khan2021sustainable}. Everything from preparing, storing, transporting, and monitoring temperature-sensitive goods from their providing areas to their receiving locations involves a sequence of activities in CSC. Food, flowers, medicines, and chemicals are examples of temperature-sensitive items because of their perishable nature \cite{li2014managing}. To increase sales and profits, companies are creating a cold chain in transportation by reducing costs, reducing delivery times, improving product quality, and offering personalized goods.

The Internet of Things (IoT) is a network that links physical items to the Internet and allows them to broadcast data to intelligently identify themselves and enable their location, tracking, monitoring, and management with the assistance of radio frequency identification tags, sensors, actuators, and positioning systems \cite{tan2010logistics}. IoT allows previously inaccessible physical and everyday items to separately gather and share data via the use of built-in sensors and network connection. IoT emerged from networked devices, which started at MIT's Auto-ID Center in the nineties, according to Center Director Kevin Ashton, who purportedly developed the concept of IoT at the turn of the millennium. The current concept of IoT is supported by the addition of GPS devices, smartphones, social networks, cloud computing, and data analytics to existing networks \cite{ben2019internet}. IoT plays a significant part in Industry 4.0, which encompasses a variety of innovative and revolutionary technologies, cyber-physical systems, IoT, and cloud computing \cite{frederico2019}. It links objects, sensors, actuators, and other smart technologies, allowing immediate access to physical commodity data to support new services with high efficiency and productivity \cite{li2015engineering}. IoT deployment may boost the likelihood of reliably transferring flow from source to sink, which is the potential of the cold chain network to withstand increased rivalry among enterprises \cite{nguyen2023estimation}.

Improved operating processes, reduced risk and expense, greater visibility and transparency, and greater adaptability and flexibility across the supply chain are merely a few of the many advantages of incorporating IoT into the supply chain \cite{mineraud2016gap}. IoT technology may drastically boost the monitoring and control of CSCs, particularly in areas such as medication, food, and logistics where temperature-sensitive commodities are moved and preserved \cite{salunkhe2016iot}.

This study conducts an exhaustive analysis of the barriers that hinder the successful implementation of the IoT in cold supply chains. Integrated interpretive structural modeling and decision-making trial and evaluation laboratory (ISM-DEMATEL) is an extensively used technique for decision-making and analysis in complex systems. ISM is beneficial for building hierarchical links among factors and analyzing the structure of complex systems, whereas DEMATEL delivers a quantitative assessment of the intensity and direction of interactions between factors \cite{george2014interpretive}\cite{du2021hierarchical}. Both approaches allow for a combination of qualitative and quantitative research, with DEMATEL specifically addressing the feedback loops. They are straightforward and obvious, making them easier to explain and grasp by decision-makers \cite{zhang2020exploring}\cite{li2019risks}\cite{alinezhad2019dematel}.

The increasing volume of international trade and consumer spending has spurred the rapid growth of supply chains for perishable commodities. The logistics for ``cold chain" items, such as perishables and frozen foods, constantly evolve. Global revenue for the sector is forecasted to reach USD 277.69 billion by 2023, growing at a CAGR (cumulative aggregate gross revenue) of 7.24 percent between 2018 and 2023 \cite{wanganoo2020real}.

\begin{figure*}[!hbtp]
\renewcommand\thefigure{1}
\centering
\includegraphics[width=1\textwidth]{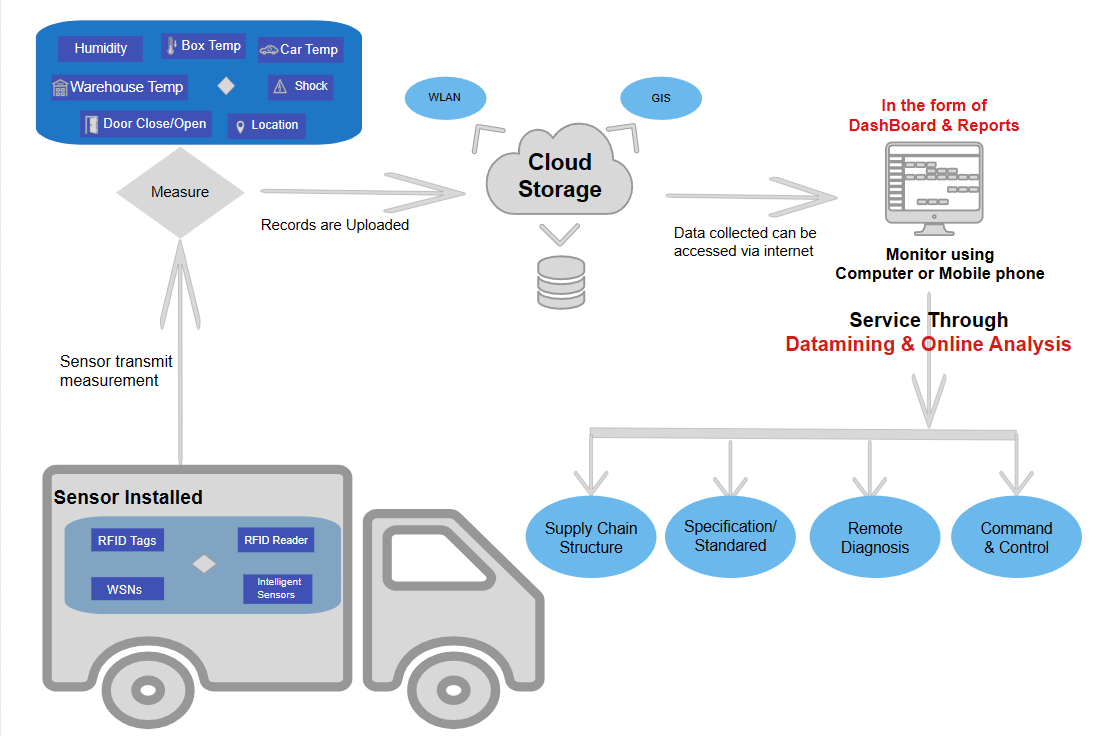}
\caption{\bf IoT implementation in the cold supply chain.}
\label{fig1}
\end{figure*}

A simplified and technology-enhanced logistics method for CSC IoT implementation is illustrated in Fig \ref{fig1}. The figure depicts IoT sensors installed across the CSC to track moisture, temperature, and other important factors, including refrigerated vehicles, warehouses, and storage facilities. The chosen parameters were controlled and monitored using a network of sensors and data-collecting microcontrollers \cite{mohammed2022design}. To provide immediate insight into the state of products in transit, data collected from these sensors are wirelessly connected and relayed in real-time to a centralized system \cite{shih2016integrating}. In sectors such as pharmaceuticals and food, where maintaining the proper temperature is crucial to product quality and safety, IoT devices make sure that temperature-sensitive commodities are stored and transported within a designated temperature range \cite{hasanat2020iot}\cite{kumar2019estimation}. By considering traffic, weather, and supply quality, IoT may help enhance delivery routes \cite{chen2020research}. This guarantees that the perishable goods arrive on schedule and in the best possible shape. IoT sensors can monitor inventory levels to minimize overstocking or understocking of temperature-sensitive products, which lowers spoilage and perishable goods losses \cite{tsang2018internet}. Well-organized logistics and transportation systems significantly affect product quality. Adopting a high-quality logistics system, especially for perishable items, involves frequent temperature monitoring, real-time product quality assessment, and shelf life determination \cite{esmizadeh2021cold}.

The objectives of the study are to apprehend the following:
\begin{enumerate}

       \item To identify the dominant barriers of IoT implementation in cold supply chains.
       
       \item To Investigate the influential relationship among barriers using ISM and DEMATEL
             technique.
        \item To identify the most significant barriers and underlying causes and effects hindering the enhancement of IoT implementation.
        
\end{enumerate}

Several studies have delved into barriers within specific countries and contexts; however, these investigations are not without their limitations. Some have explored sectors unrelated to the CSC, and while certain studies have focused on IoT adoption in this context, they often fall short of establishing connections among identified barriers or evaluating their respective effects on the overall system. Moreover, the exploration of IoT implementation barriers in CSCs represents a significant gap in the literature, with no prior comprehensive studies identified in this specific domain. The lack of scholarly attention to this crucial aspect of CSC underscores the need for an in-depth investigation. Our research is a pioneering effort to bridge this gap by employing an integrated ISM-DMATEL approach.

The literature gap is particularly evident in the scarcity of empirical examinations and theoretical frameworks dedicated to understanding the intricacies of IoT implementation in CSCs. The unique challenges posed by the CSC environment, characterized by stringent temperature control requirements and perishable goods logistics, warrant further exploration.

To address these gaps, this study contributes to the literature in the following ways:

\begin{enumerate}

    \item An in-depth analysis of research on essential barriers to CSC IoT adoption, considering various perspectives, including supply chain stages, environmental effects, socio-economic scenery, and data complexity.

    \item A survey-based approach was employed to gather data and expert opinions regarding these barriers.

    \item The utilization of the ISM method to determine the interdependencies among the selected barriers.

    \item The MICMAC strategy was applied to comprehensively analyze the fundamental constructs, thereby offering valuable insights into their significance.

    \item The DEMATEL approach was used to validate causal relationships among the components within the model.

    \item Key barriers that play a fundamental role in efficiently managing the cold chain are highlighted, offering insights to overcome challenges in IoT adoption.

    \item Systematic identification and ranking of barriers based on their influence on the system addresses the current lack of comprehensive assessments in the literature.

    \item Contribution towards filling research gaps by focusing specifically on the CSC, distinguishing it from studies that extend across different sectors.

    \item Providing a structured approach to understanding and addressing the challenges associated with IoT implementation in CSC, contributing to enhanced operational efficiency and reliability.

    \item It offers valuable insights for enterprises, governments, and technology developers to navigate and overcome obstacles in integrating IoT into CSC operations.

\end{enumerate}

The rest of the paper is arranged as follows: ``\hyperref[sec2]{Literature review}" provides a comprehensive literature review of IoT implementation in CSCs, including implementation challenges and comparable methodologies utilized in the present era; ``\hyperref[sec3]{Methodology}" presents the research methodology that was used for the study; ``\hyperref[sec4]{Numerical illustration for results interpretation}" covers the data analysis, along with the application of ISM-DEMATEL to the obtained barriers; ``\hyperref[sec5]{Discussions}" describes the results of the implementation of ISM-DEMATEL techniques and discusses the results. Finally, the study concludes with conclusions, limitations, and recommendations for future work in ``\hyperref[sec6]{Conclusions and future works}."

\section{Literature review}
\label{sec2}
The aim of this section is to provide a summary of how CSC procedures may use IoT. We used Google Scholar and many other publishers' websites, including Springer, Elsevier, Emerald, and Taylor-and-Francis, to compile the literature most relevant to our study.

\subsection{Internet of Things (IoT)}
The IoT is among the most revolutionary technological developments brought about by the digital era \cite{hasan2022revolutionizing}. The Internet of Things is anticipated to become the most popular form of internet traffic in the next years, with a predicted CAGR of 40 in the commercial sector \cite{zhou2016guest}. The economy, communication patterns, and technology development are all expected to be profoundly affected \cite{okoroguinternet}.
First invented in 1999 by Kevin Ashton and many other scholars at the Massachusetts Institute of Technology, IoT has gained broad interest because it can provide comfort, efficiency, and competitive benefits to the corporate sector \cite{aryal2020emerging}. The IoT is an essential idea in the coming period. Numerous smart buildings and smart houses are built using IoT technologies \cite{shi2020framework}. The IoT presents a new opportunity to reduce risks, manage complexity, and produce concrete economic benefits by increasing transparency and flexibility across the supply chain \cite{al2020achieving}. The physical components that link IoT include wireless devices such as portable computers, smartphones, wearable gadgets, RFID tags, RFID readers, and wireless sensors. Users may interact with the IoT network using a variety of devices, including cellphones, implanted RFID transponders, wirelessly connected PCs, and wearable technology \cite{krotov2017internet}. The popularity and success of IoT have skyrocketed in recent years. Industry leaders, technologists, and consumers recognize the importance of IoT, which has led to an industrial revolution and automated operations in organizations and homes \cite{saleem2018iot}. Human-to-thing communication and autonomous coordination among ``things" is possible even when they are kept at a facility or transported from one entity in the supply chain to another. Increased supply chain visibility, responsiveness, and adaptability to a variety of SCM challenges are all made possible by IoT \cite{ben2020role} \cite{ben2019internet}.

\subsection{IoT in CSC}
IoT, often known as the third wave of IT, has revolutionized supply chain management (SCM) and other sectors. This helps expand the supply chain function by offering technical assistance for greater visibility, stability, and intelligent management \cite{cui2018supply} \cite{sallam2023internet}. Ben-Daya \textit{et al.}  \cite{ben2019internet} analyzed the existing literature to learn more about the IoT and how it influences SCM. IoT has transformed supply chain management by improving efficiency, allowing predictive maintenance, and enabling real-time reactions to change circumstances, improving operational efficiency, waste reduction, and overall customer happiness \cite{yesodha2023iot}\cite{yesodha2023iot}. Bahl \cite{bahl2018role} explores how supply chain management has been transformed by IoT technology, which offers real-time data insights, improves visibility, and allows proactive decision-making to satisfy consumer expectations in a changing marketplace and optimize operations. In addition, the majority of research has concentrated on the steps in the supply chain pertaining to food production, distribution, and processing. Kamble \textit{et al.} \cite{kamble2019modeling} proposed an integrated ISM-DEMATEL model to identify the most significant barriers to the widespread deployment of IoT technologies in the food retail supply chain. The two most significant barriers to the widespread use of IoT were a lack of government regulation and insufficient network infrastructure. Nguyen \textit{et al.} evaluated the reliability of a cold chain network by considering both flow and time factors simultaneously, with the ultimate goal of determining the network reliability for actual products. Shashi \cite{shashi2022digitalization} revealed two primary themes: the existing cold chain system's known or unknown limits, as well as the use of digital enablers based on the IoT. 

Numerous deployment strategies, constraints, and barriers in cold chain monitoring applications have been discussed \cite{badia2018new}.
Sallam \textit{et al.} \cite{sallam2023internet} analyzed future developments in the field of supply chain management and offers a thorough analysis of the IoT applications in this area, including problems, possibilities, and best practices.
The feasibility of implementing modern technologies such as IoT to enhance cold SC was assessed by Gupta \textit{et al.} \cite{gupta2019modeling}. To enhance service quality, the organization must invest additional funds for maintenance. Umamaheswari \textit{et al.}
\cite{umamaheswari2020iot} suggested smart cold storage that uses cutting-edge supply chain technology and the IoT to boost productivity and reduce wait times at all levels. Luo \textit{et al.} \cite{luo2016intelligent} presented a smart monitoring system that stores data on defined servers to monitor cold chain items in real time while cutting costs, enhancing network capacity, and simplifying protocols. Cold chain transit is easy to track using this strategy.
Afreen and Bajwa \cite{afreen2021iot} implemented a system that utilizes the IoT to track and notify users of changes in cold storage conditions such as humidity, temperature, light, and gas concentration in real-time., alerting workers when these variables exceed safe thresholds. Goodarzian \textit{et al.}  \cite{goodarzian2023designing} proposed a model for improving the precision, swiftness, and fairness of vaccine injection and implemented an inaugural responsive green-cold vaccine supply chain network in response to the COVID-19 pandemic. Intelligent control technology, cellular networks, wireless communication technology, temperature and humidity sensors, IoT, and database technology all come together to form a cold chain management system\cite{zhang2019nb}. Ceken and Abdurahman \cite{cceken2019simulation} presented an IoT-based cold chain logistics system that controls and monitors the cold chain's ambient temperature in real time and estimates perishable item shelf life to help all stakeholders make better decisions.
The paper by Tsang \textit{et al.} \cite{tsang2018internet} presented an IoT-based risk monitoring solution to handle product quality and safety concerns in the cold chain. Using this technology, the cold chain can be monitored in real-time, and risks can be assessed to ensure the safety of workers. Hernandez and Yamaura \cite{hernandez2017cold} suggested that IoT-enabled devices may reduce pharmaceutical waste caused by temperature changes. It also demonstrated consistency in IoT implementation, including infrastructure, security, and device management. Owing to perishable food and cross-regional transit, cold chain systems encounter several challenges in the perishable food supply chains. Atyam \textit{et al.} explained the complexity and problems associated with modern supply chain management, the importance of IoT in monitoring items, and how IoT and blockchain might improve supply chain performance. Kumar \textit{et al.} \cite{kumar2023visualization} highlighted the need of efficient cost management and waste reduction as key performance indicators and also emphasized the possibility of boosting operational efficiency and optimizing resource utilization via the integration of the IoT. IoT-enabled smart monitoring and control systems could provide long-term solutions to energy consumption and managerial reluctance in the cold supply chain \cite{kumar2023sustainable}.  This study of Zhang \textit{et al.} \cite{zhang2017modeling} addressed these difficulties by creating a perishable food supply chain model utilizing real-time IoT data and two supply hubs to increase performance. Future cold supply networks can become more resilient and sustainable if IoT technologies are used. This advancement is expected to result in less waste and decreased operational expenses, therefore benefiting both businesses and consumers \cite{gupta2019modeling} \cite{abdel2018internet}.

\subsection{IoT implementation barriers}

IoT deployment is challenging because of device and data security, interoperability issues, scalability, trust, privacy concerns, financial issues, lack of qualified personnel, and compliance with data privacy laws. First, it is necessary to identify the barriers that hinder IoT implementation in CSC in order to overcome these challenges. \autoref{tab1} lists the barriers to implementing the IoT in CSC.

\begin{table*}[!ht]
\renewcommand\thetable{1}
\centering
\caption{\bf Barriers to IoT implementation in the cold supply chain}
\label{tab1}
\resizebox{\textwidth}{!}{
\begin{tabular}{lm{3cm}m{10cm}m{4cm}}
\hline
\textbf{Sl No.} & \textbf{Barriers}                     & \textbf{Implied Meaning} & \textbf{References} \\ \hline
01 & Regulatory compliance &
The CSC has strict temperature and data management guidelines. It may be challenging to implement IoT technology and meet these requirements. In the pharmaceutical and food industries, strict regulations are tough to follow. Failing to satisfy these regulations may result in product recalls or legal concerns. & \cite{babagolzadeh2020sustainable} \cite{stuurman2016iot} \cite{jerkins2017motivating}  \cite{zhou2015iot} \cite{kumar2023visualization} \\

02 & High implementation cost &
In order to adopt the IoT, it may be necessary to make a sizable initial investment in things like sensors, devices, and infrastructure. Transporting and storing cold requires plenty of energy. Costs may rise with power prices.& \cite{urbano2020cost} \cite{badia2018new} \cite{gupta2019modeling} \cite{abdel2018internet} \\

03 & Lack of skilled workforce & 
Multiple sensors, data analytics, and system integration make IoT-enabled CSCs difficult. Implementing, deploying, and administering these systems may be challenging without trained people. Integration of IoT into CSCs demands expertise. Without qualified workers, organizations may struggle to analyze this data.& \cite{manavalan2019review} \cite{madanayake2020investigating} \cite{hussain2017internet} \cite{seet2021iot} \\

04 & Data security and privacy &
Data generated and sent by IoT devices is often private. Extremely regulated sectors, such as healthcare and pharmaceuticals, make it difficult to safeguard this data and guarantee privacy compliance.& \cite{zhang2020blockchain} \cite{nakamura2019tboi} \cite{hernandez2017cold} \\

05 & Technological infrastructure & 
Integrating the IoT into an already established supply chain might be challenging. On the other hand, it's likely that some of the more recent IoT technologies won't be compatible with more established networks and software. IoT devices can't be deployed and used effectively without the right infrastructure in place.&  \cite{luo2016intelligent} \cite{mohsin2017iot} \cite{hernandez2017cold} \cite{karimi2022developing} \\

06 & Scalability challenges &
Scalability is a challenge for numerous sectors. It may be difficult to implement IoT throughout a complex supply chain. Cold supply networks need strict data monitoring and traceability. Implementing and running these systems need accurate data. Coordinating many linked devices and data becomes more difficult as IoT networks grow.&  \cite{raut2019improvement} \cite{mohsin2017iot} \cite{li2021development} \cite{hussain2017internet} \\

07 & Cultural and ethical consideration & 
Cultural norms and expectations must be addressed when adopting IoT devices that monitor behaviors and data. IoT devices that monitor and enhance work processes may affect employees differently by culture. Technological developments may not benefit all workers. As IoT automation reduces human control, workers may fear they will be outdated. &  \cite{sodhi2021supply} \cite{stuurman2016iot} \cite{seet2021iot} \\
08 & Acceptance and adoption & 
IoT implementation may involve organizational training and cultural changes; therefore, resistance to change and acceptance may be a significant obstacle. Several parties in the CSC may not understand IoT's potential benefits, generating resistance. Worker resistance to the current system may hinder adoption. & \cite{kokkinos2018efficient} \cite{gillespie2023real} \\
09 & Data integration and management issue &
Constructing a cohesive and efficient system might be difficult due to possible communication difficulties across different IoT devices and platforms. The IoT is the source of a lot of data. Without proper data management and analytics tools, it is very challenging to deal with this massive amount of data, let alone analyze it and derive any meaningful conclusions from it. & 
\cite{luo2016intelligent} \cite{sarkar2022iot} \cite{wanganoo2020real} \cite{li2021development} \\
10 & Reliability issue & 
Inefficient asset monitoring and network concerns may cause asset loss, theft, customer distrust, delays, data loss, and inefficiency. IoT device failures may interrupt operations and raise maintenance expenses. Improper data input or modification without validation might affect reliability, resulting in non-compliance with industry standards and legal issues. & \cite{badia2018new} \cite{tsang2018internet} \cite{zhou2015iot} \cite{nakamura2019tboi} \\
11 & ROI uncertainty & 
A thorough cost-benefit analysis is required to determine the potential return on investment. The overall cost of an IoT deployment might be more than expected if the time it takes to see a return on investment is greater than expected. & \cite{badia2018new} \cite{gupta2019modeling} \cite{kumar2023sustainable} \\
12 & Extreme environment issue &
Extreme weather is a common occurrence for cold supply networks. Radio wave frequency is impacted by nearby elements, including metals and water, resulting in variable performance. It might be difficult to maintain the functionality and reliability of IoT devices when they are in cold storage or in transit. & \cite{tsang2017iot} \cite{badia2018new} \cite{afreen2021iot} \cite{shashi2022digitalization} \\
13 & Cold chain network &
Powering remote or mobile supply chain equipment might be difficult. IoT devices may have trouble delivering data in rural or poor areas with poor network coverage. Many temperature-sensitive items can not resist slight temperature fluctuations during transport, storage, and handling, which could affect the cold chain network. & \cite{soh2021cold} \cite{zhang2021pharmaceutical} \cite{wanganoo2021nb} \cite{gillespie2023real} \cite{shashi2022digitalization} \\
\hline
\end{tabular}
}
\end{table*}

\subsection{Methodology related literature}

ISM is a computer-based learning method that assists individuals or groups in creating a visual representation of intricate connections between several aspects of a complex situation. The fundamental idea is to use the practical skills and knowledge of specialists to dissect a complex system into many sub-systems and construct a multi-level structural model\cite{george2014interpretive}. Raut \textit{et al.} \cite{raut2017identify} employed the ISM approach to establish the interconnections among drivers, enabling a comprehensive knowledge of the relative connections among the critical success factors and defining their dependency when implementing sustainability.

The DEMATEL technique has been used extensively to identify critical barriers in various vital processes, which were created to support methodical evaluations and organized decision-making processes. The effect connections map shows the relational linkages, with numbers denoting the degree of influence and arrows denoting the direction of influence. It is a comprehensive tool for assessing the interrelations between system variables and emphasizing the core-driving components \cite{du2021hierarchical}. Manoharan \textit{et al.} \cite{manoharan2022contextual} presented the DEMATEL model, which aids in comprehending the importance of drivers and barriers, creating matrices or digraphs to represent complex causal interactions, and identifying the relationships between these factors. The barrier analysis techniques employed in recent studies are presented in \autoref{simmedtble}.

\begin{table*}[!ht]
\renewcommand\thetable{2}
\caption{\bf Various barrier analysis methodologies used in recent approaches}
\resizebox{\textwidth}{!}{
\bgroup
\def\arraystretch{2.5}%
\begin{tabular}{lm{2cm}lm{6cm}m{6cm}}
\hline
\textbf{Sl no.} & \textbf{Authors}                                                                & \textbf{Methodology}                  & \textbf{Problems}                                                                                                                    & \textbf{Findings}                                                                                                                                                                                    \\ \hline
1     & \cite{gokarn2021modeling} & ISM   MICMAC                 & Figured out what matters most for reducing food waste and loss in supply chains for fresh produce                & Regulatory bodies, food policy, and market infrastructure are the most important elements that have a strong propensity to affect the lowering of FLW in FPSCs.                         \\ \hline
2     & \cite{singh2020integrated}    & DEMATEL-MMDE-ISM             & Identified and evaluated potential barriers to IoT deployment in the manufacturing industry & Highlighted the key challenges where academics and practitioners may concentrate their strategic efforts to overcome implicit concerns while deploying IoT Techniques in manufacturing. \\ \hline
3     & \cite{cui2021internet}                                                           & Bayesian   Best-Worst Method & Discovered barriers to implementing the circular economy in the manufacturing sector.                                        & Recommended a more realistic PFS performance, allowing several uses.                                                                                                                          \\ \hline
4     & \cite{abkenar2022determining}                                                      & Bayesian   Best-Worst Method & Identified multiple barriers to implementing the IoT in the food industry                       & Major barrier to the IoT in the food industry is the absence of an adequate internet infrastructure.                                                                                                                   \\ \hline
5     & \cite{kumar2018analysis}                                                     & ISM-DEMATEL                  & Examined the identified obstacles that impede the implementation of electronic waste management practices in India.                             & Key challenges pertaining to e-waste recycling are a lack of public knowledge and rules about the matter.   \\ \hline                                 
6   & \cite{meng2022study}  & DEMATEL–ISM & Identified what circumstances affected the flight crew's TSA protocol & Different levels of hierarchy, mutual impact, and the cause-and-effect link are shown for the flight crew's TSA creation and management.
\\ \hline
7 & \cite{song2020analyzing} & Fuzzy-ISM-DEMATEL &  Identified barriers to implementing sustainable purchasing practices over the internet & Improved DEMATEL, ISM, and rough set theory into a unified framework. 
\\ \hline
8 & \cite{chauhan2018interpretive} & ISM-DEMATEL & Analysed the challenges associated with waste recycling in India & For India's trash recycling infrastructure to grow, significant barriers such as a lack of funds, input material, and subsidies have to be overcome.
\\ \hline
\end{tabular}
\label{simmedtble}
\egroup
}
\end{table*}

The literature investigation discusses every element of the CSC, covering transportation, cold storage, procurement, and occupational security across the cold chain. IoT supports SCM by enhancing visibility and stability, but problems include regulatory limitations and poor infrastructure. Most studies focus on distinct problems and seek to solve these challenges using the solutions provided. In the combined ISM-MICMAC-DEMATEL approach, ISM-MICMAC is used to construct hierarchical relationships among barriers and analyze the structure of systems, while DEMATEL delivers a numerical evaluation of the strength and direction of interactions between factors. Numerous barrier analysis strategies have been employed in modern methods, which are also used to analyze obstacles. The choice between these tactics varies depending on individual circumstances, available data, and decision-makers' preferences. Thirteen barriers, and from the different barrier analysis techniques utilized in recent approaches, ISM-DEMATEL was chosen to analyze these barriers.

\section{Methodology}
\label{sec3}
Research initiation involves several key stages, including the design of surveys, meticulous data collection, selection procedures, and rigorous reliability testing, as illustrated in Fig \ref{fig2}. Subsequently, the second phase unfolds with a process flow diagram outlining the development of the ISM model. The third phase is characterized by a process flow diagram that delineates the systematic steps employed in constructing the DEMATEL model.

\begin{figure*}[!hbtp]
\renewcommand\thefigure{2}
\centering
\includegraphics[width=1\textwidth, height=1.5\textwidth]{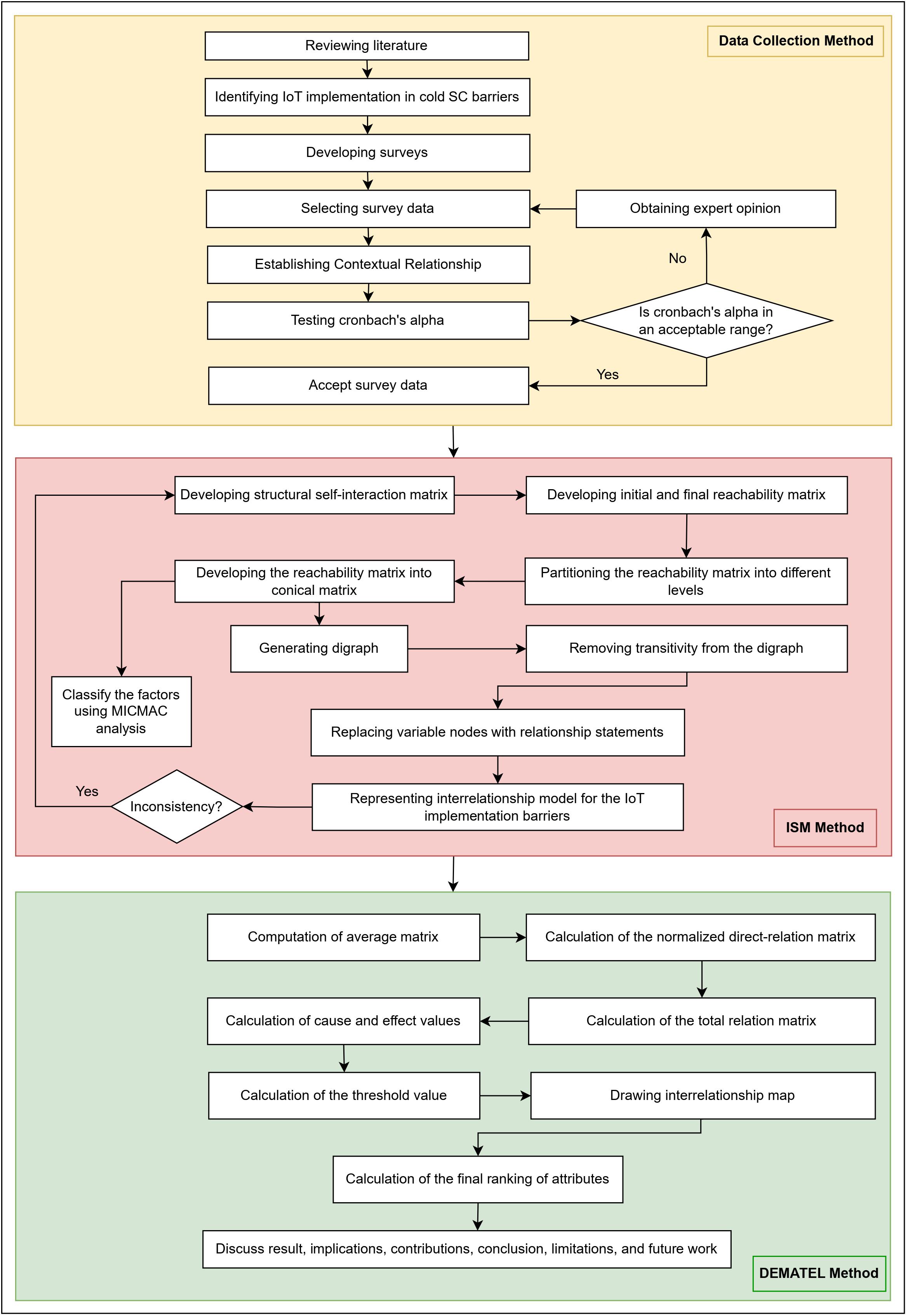}
\caption{\bf Flowchart depicting the research methodology, encompassing phases such as data collection and validation, implementation of ISM-MICMAC, and execution of the DEMATEL procedure.}
\label{fig2}
\end{figure*}

\subsection{Data collection method}

\subsubsection{Barriers identification}

This involves identifying and recognizing the possible obstacles or challenges inside the system. These obstacles may be technical, organizational, or connected to other system operation elements. Using literature research and expert comments, potential barriers to the system were identified, as described in Table \ref{tab1}.

\subsubsection{Survey development and data selection procedure}

Thirty experts participated in the survey. The diagonal values in the ISM method are zero. Only one side of the diagonal value is used to create the initial reachability matrix. All 13 rows and 13 columns of the matrix were surveyed, with the exception of diagonal values. This verification process guaranteed that the data provided by the respondents were valid. After determining that the impact of barrier 1 on barrier 2 is V, an expert must conclude that the impact of barrier 2 on barrier 1 is A. If O and X are diagonal, their opposing sides must be the same. These checks will improve the quality of the survey, even if the initial reachability matrix ultimately determines the entire system. 

The DEMATEL impact factor survey, which is likewise associated with the ISM survey, is also measured on a scale from 0 to 4. The next issue concerns the impact factor, which must be greater than or equal to 1 if the effect of barrier 1 on barrier 2 is V. A value of 0 is required for the influence factor if it is O. A positive number is also required for X. Therefore, this cross-checking method will enhance the quality of the survey.

Twelve reasonable answers were selected for each question, and twelve $13\times13$ matrices were constructed. The average of these matrices was calculated to determine the average matrix. For the ISM, the structural self-interaction matrix is also prepared using the value that appears most often in these logical replies.

Table \ref{demographic} provides a brief overview of the various characteristics of the expert panel. Quantification of each category offers vital information regarding the makeup of the group. A table of this type assists in visually representing and examining the distribution of expertise. The organized survey data from these 30 experts helps the comprehensibility of the information, which is essential for generating the outcome. The sex distribution comprised 26 males and 4 females. Thirteen panelists are below 30, 8 fall in the 30–40 range, 5 are aged 41–50, and 4 are 51 years and above. Regarding experience in their respective fields, 10 experts had less than five years, 7 possessed 5–10 years, 10 had 10–15 years, and 6 held 15 years and above. The domain of expertise includes eight in academics (Industrial Engineering), five in System Integration, four in Data Analysts, six in Data Security, and seven in cold chain executives.

The survey was conducted using Google Forms, and the recruitment period was from 27 Sep 2023 to 11 Oct 2023. Participants were selected based on their expertise in supply chain, system integration, data analysis, data security, and cold chain. The survey consisted of 312 questions, focusing on the influential relationships among the barriers. Confidentiality of participant responses was strictly maintained. All the collected data were anonymized and stored. Only the research team had access to the raw data, and the findings were reported in aggregate form to ensure the anonymity of the participants. This study adhered to the ethical principles outlined in the Declaration of Helsinki. Participants were provided with informed consent forms detailing the purpose of the study, their rights as participants, and the procedures for data handling. Participation in the survey was voluntary, and participants had the right to withdraw at any time without penalties. This study was approved by the Ethics Review Board of the Research \& Extension Center, Khulna University of Engineering \& Technology, Bangladesh, to conduct this survey research.

\begin{table}
\centering
\caption{\bf Demographic profile of the selected experts}
\begin{tabular}{lc} 
\hline
\textbf{Demographics} & \textbf{Number of experts} \\ 
\hline
\textit{\textbf{Gender}} & \textit{\textbf{~}} \\
Male & 26 \\
Female & 4 \\
~ & ~ \\
\textit{\textbf{Age}} & \textit{\textbf{~}} \\
Below 30
  years & 13 \\
30–40
  years & 8 \\
41–50
  years & 5 \\
51 years
  and above & 4 \\
~ & ~ \\
\textbf{\textit{Experience in their field}} & \textbf{\textit{~}} \\
Less
  than five years & 10 \\
5–10
  years & 7 \\
10–15
  years & 7 \\
15 years
  and above & 6 \\
~ & ~ \\
\textbf{\textit{Domain of expertise}} & \textbf{\textit{~}} \\
Academics
  (Industrial Engineering) & 8 \\
System
  Integration & 5 \\
Data
  Analyst & 4 \\
Data
  Security & 6 \\
Cold
  chain executives & 7 \\
\hline
\end{tabular}
\label{demographic}
\end{table}

\subsubsection{Establish relationships between barriers}

Once possible obstacles are discovered, it is essential to understand how these barriers are interrelated or impact each other. This step helps in understanding the complexity of the system and how different challenges may be connected. To determine which pairs of barriers require further investigation, a contextual relationship between them is established.

\subsubsection{Cronbach's alpha testing}

The reliability coefficient Cronbach's alpha can be used to evaluate the internal consistency of the survey or test items. Many psychological and educational surveys and assessments have used Cronbach's alpha for their formulation and testing. This aids researchers and practitioners in ensuring that their measuring equipment accurately measures the idea in question. Cronbach's alpha reliability is a measure of how well a sum or average of k measurements holds, where k measures might be anything from k raters to alternate forms to items on a questionnaire or exam. We used a 156-item questionnaire (k = 156) to measure implementation barrier influence, with each item scored on a 0–4 Likert scale. A Cronbach's alpha value should exceed 0.7 since a value equal to or more than 0.7 falls within an acceptable range \cite{bonett2015cronbach}.

The variance of the total score (270.52) and the sum of the item variances (67.88) provide this dependability measure. The survey dataset had an acceptable degree of internal consistency, as indicated by a Cronbach's alpha value of 0.7534. This measuring tool reliably captures the targeted construct because the coefficient is greater than the generally recognized requirement of 0.7.

\subsection{Interpretative structural model (ISM)}\label{sec:3.1}
The initial development of ISM can be attributed to J. Warfield in 1973 who examined intricate socioeconomic systems  \cite{chauhan2018interpretive}. This method may help individuals or teams better grasp a topic by organizing their understanding of it into a network diagram. The following is a detailed description of how ISM works \cite{zhang2020exploring}\cite{li2019risks}.

\subsubsection{Step 1: Developing structural self-interaction matrix (SSIM)}
\label{ISMstep3}
The pair-wise connection between barriers in the system can be observed by employing SSIM.
The following four symbols indicate a unique set of relationships:
\begin{itemize}
    \item V: Barrier $i$ influences barrier $j$;
    \item A: Barrier $j$ influences barrier $i$;
    \item X: Barriers $i$ and $j$ influence each other;
    \item O: Barriers $i$ and $j$ have no relationship.
\end{itemize}

\subsubsection{Step 2: Developing reachability matrix}
\label{ISMstep4}
The SSIM generates a reachability matrix; four symbols are converted to 0s and 1s to implement the SSIM in the initial reachability matrix.
matrix.

Specific rules of replacement are as follows:
\begin{itemize}
    \item If $D(i,j)$ of SSIM is shown with the symbol ``V," $D(i,j)=1$ and $D(j,i)=0$ should be assigned in the initial reachability matrix;
    \item If $D(i,j)$ of SSIM is shown with the symbol ``A," $D(i,j)=0$ and $D(j,i)=1$ should be assigned in the initial reachability matrix;
    \item If $D(i,j)$ of SSIM is represented by an ``X," then $D(i,j)=1$ and $D(j,i)=1$ should be placed in the initial reachability matrix,
    \item If $D(i,j)$ of SSIM is represented by an ``O," then $D(i,j)=0$ and $D(j,i)=0$ should be assigned.
\end{itemize}

\subsubsection{Step 3: Developing final reachability matrix}
A transitivity test is then performed on the initial reachability matrix. ISM presupposes, as a given, that contextual relations are transitive. If barrier A is connected to barrier B, and B is related to barrier C, then barrier A must also be related to barrier C.

\subsubsection{Step 4: Developing level partition}
Level partitioning refers to categorizing or classifying distinct levels inside the information system. This involves dividing the system into levels or hierarchies to better understand its structure at various levels. The resultant final reachability matrix is segmented into several hierarchical levels.

\subsubsection{Step 5: Developing conical form}
The conical form describes structuring or organizing an information system in an arrangement that resembles a conical shape, indicating a hierarchical structure. Subsequently, a conical form of the reachability matrix was created.

\subsubsection{Step 6: Developing ISM model}
A directed graph was constructed using a reachability matrix to eliminate transitive linkages. By exchanging the nodes of the variables for statements, they converted into an ISM from the resultant digraph.

\subsubsection{Step 7: Reviewing to check for inconsistency}
After establishing the level partition, conical shape, and ISM model, it's important to examine the full system to detect and resolve discrepancies. This step verifies that the system was consistent and functions smoothly without conflicting factors. The generated ISM model is reviewed for conceptual coherence, and any necessary revisions were implemented.

\subsection{MICMAC analysis}
The multiplication of impact matrix applied to classification (MICMAC) study seeks to create a graph to categorize various barriers. The MICMAC algorithm is based on the multiplication characteristics of matrices \cite{dewangan2015enablers}.

The ISM framework requires the construction of a conical matrix to depict the interrelationships of a system's components. For each element, the driving and dependence powers were calculated by adding the values in the rows and columns, respectively. These elements were further classified according to the effectiveness of their driving and dependent power in the MICMAC analysis. Factors were grouped into four categories: autonomous, dependent, linkage, and independent \cite{sindhwani2017modelling}. 

\begin{enumerate}
    \item Group I (Autonomous factors): Comprises autonomous factors characterized by minimal interdependencies.
    \item Group II (Dependent factors): Encompasses dependent factors exhibiting a substantial reliance but limited generative capacity.
    \item Group III (Linkage factors): Consists of linkage factors possessing a robust propelling force and reliability.
    \item Group IV (Independent factors): Involves factors that significantly impact the outcomes of sustainable energy initiatives but are not significantly influenced by other variables. These factors can be utilized to manipulate other elements, thereby contributing to the preservation of sustainable energy objectives. 
\end{enumerate}

\subsection{DEMATEL}
The DEMATEL method transforms the complex web of interrelations between barriers into a clear structural model of the system by categorizing them into two groups: causes and effects. This is a practical method for ranking the relative importance of different aspects of a complex system to inform strategic decisions over the long run and identify potential development areas \cite{si2018dematel}. It is extensively implemented across numerous domains, such as SCM, traceability, smart cities, healthcare, and consumer behavior \cite{khan2022exploration}. The following is a detailed description of how DEMATEL works
\cite{alinezhad2019dematel} \cite{kamble2019modeling}.

\subsubsection{Step 1: Computation of average matrix}

The expert panel that filled out the survey was asked to provide an integer value between 0 and 4 to represent the degree of influence they saw between any two barriers, using five different integer scales as “very high influence (4)”, “high influence (3)”, “moderate influence (2)”, “low influence (1)”, and “no influence (0)”.
A respondent's estimate of how much variable $i$ influences variable $j$ is denoted by the $x_{ij}$ symbol.
For $i=j$, 0 is assigned to each diagonal element.

For every respondent, we may generate a non-negative $n \times n$ matrix using the formula $ X^{k}= [x_{ij}^{k}] $, where $k$ means that there were a total of responders with $1\leq k\leq H$ and the total number of factors is denoted by $n$. In this case, the $H$ responder matrices are denoted by $X^{1}, X^{2}, X^{3},..., X^{H}$.

The average matrix can be created as
\begin{equation}
a_{ij}= \frac{1}{H}\sum_{H}^{k=1}x_{ij}^{k}
\label{eqn3.1}
\end{equation}

\subsubsection{Step 2: Calculation of the normalized direct-relation matrix}
Normalize direct-relation matrix $D$ by $D = A \times S$, where
\begin{equation}
S=max_{ij}\left \{ {\frac{1}{max_{i}\sum_{j=1}^{n}\left | x_{ij} \right |}},{\frac{1}{max_{j}\sum_{j=1}^{n}\left | x_{ij} \right |}}\right \}
\label{eqn3.2}
\end{equation}

\subsubsection{Step 3: Calculation of the total relation matrix}

The total relation matrix $T$ is represented as 
\begin{equation}
T=D(I-D)^{-1},
\label{eqn3.3}
\end{equation}
where $I$ is the identity matrix.

\subsubsection{Step 4: Calculation of cause and effect values}

$R$ and $C$ values, representing the sum of the values in the rows and columns, are now used in the calculations. Where,
\begin{equation}
R=\left [ R_{i} \right ]_{n\times 1}=\left [ \sum_{n}^{j-1}t_{ij} \right ]_{n\times 1} ; i=1,......,n
\label{eqn3.4}
\end{equation}

\begin{equation}
C=\left [ C_{j} \right ]_{1\times n}^{'}=\left [ \sum_{n}^{j-1}t_{ij} \right ]_{1\times n}^{'} ; j=1,....,n
\label{eqn3.5}
\end{equation}

The transpose value of the $j^{th}$ column, denoted by $\left [ C_{j} \right ]^{'}$, and the sum of the values of the $j^{th}$ column, $C_{j}$, shows the effect of the $j^{th}$ attribute on the other attributes. The influence of the $i^{th}$ row, denoted by $r_i$, indicates the cause of the $i^{th}$ attribute on other attributes. The sum $(R_i+C_i)$ represents the cumulative effects experienced and exerted by factor $i$. This signifies the degree of importance of element $i$ in the overall system. The difference $(R_i-C_i)$ indicates the net effects of factor $i$ on the system. If the difference between $D_i$ and $R_i$ is positive, the influence factor $i$ may be considered as a net cause. Conversely, if the difference between $D_i$ and $R_i$ is negative, then factor $i$ can be regarded as a net receiver.

\subsubsection{Step 5: Calculation of the threshold value}
Determination of the threshold value is necessary to construct an interrelationship map. The interrelationship map only displays equations with penetration matrix values greater than the threshold value. The mean of the elements inside the entire penetration matrix is computed using \autoref{eqn3.6}.

\begin{equation}
a= \frac{\sum_{i=1}^{n}\sum_{j=1}^{n}t_{ij}}{N}
\label{eqn3.6}
\end{equation}

\subsubsection{Step 6: Drawing interrelationship map}
\label{DEMATELstep6}
An interrelationship map is created to comprehensively analyze the final solution, considering the values of $R_i+C_i$ and $R_i-C_i$ for each attribute in relation to the threshold value.

\subsubsection{Step 7: Calculation of the final ranking of attributes}
\label{DEMATELstep7}
The interrelationship map derives a potential framework and hierarchy of these factors. This is achieved by organizing the values of $R_i+C_i$ in descending order.

\section{Numerical illustration for results interpretation}
\label{sec4}

\subsection{ISM-MICMAC application}
\label{sub:ism}
ISM and MICMAC techniques are strong ways to create a hierarchy and reveal contextual interactions between IoT adoption barriers, especially regarding the CSC. The principal aim of utilizing MICMAC is to evaluate the importance of the dimensions that affect a system. Using this methodology, researchers can examine the complex interrelationships and dependencies between different components and identify important factors that have a major impact on the system as a whole. The ISM-MICMAC technique has two main phases that have been applied in various settings such as supply chains, industry, and healthcare. A hierarchical structure of the chosen IoT adoption barriers was first constructed using the ISM. The relative importance and hierarchy of each barrier were then determined by methodically examining cause-and-effect interactions. After the dimensions are established, professionals evaluate the correlations, resulting in the development of a binary relation matrix that is then converted into a reachability matrix. The hierarchy of the barriers is displayed in this matrix, highlighting their key and consequential functions within the CSC framework. The MICMAC technique is used in the next step to further explore the relationships and interactions among the various IoT adoption barriers that have been identified. The emphasis is on understanding the dynamic behavior of barriers and classifying them according to their reliance and driving forces. Identifying driving barriers with high influence but low dependence and dependent barriers with high influence but low driving power is made easier using MICMAC analysis. With this categorization, the role of each barrier in the bigger picture and the vital links between them can be better understood.

\subsubsection{Structural self‑interaction matrix (SSIM) development}
Experts' judgments on the relationships between the 13 identified barriers are used to inform the SSIM's development, which involves using four symbols to depict these interactions in a pairwise fashion, as described in \autoref{ISMstep3}. In Table \ref{tab:table 4.1.1}, SSIM depicts the element intrinsic linkages and internal dynamics.

\begin{table*}[!ht]
\renewcommand\thetable{4}
\centering
\caption{\bf Structural self-interaction matrix (SSIM)}
\resizebox{.9\textwidth}{!}{
\begin{tabular}{|clccccccccccccc|}
\hline
\textbf{Sl No.} &\textbf{Barriers}                             & \textbf{1} & \textbf{2} & \textbf{3} & \textbf{4} & \textbf{5} & \textbf{6} & \textbf{7} & \textbf{8} & \textbf{9} & \textbf{10} & \textbf{11} & \textbf{12} & \textbf{13} \\ \hline
1&Regulatory Compliance                 &   & V & V & V & V & V & V & V & V & V  & V  & V  & V  \\ \hline
2&High implementation cost              &   &   & A & A & A & A & O & V & A & V  & X  & A  & A  \\ \hline
3&Lack of skilled workforce             &   &   &   & A & A & V & X & V & X & V  & V  & O  & A  \\ \hline
4&Data security and privacy             &   &   &   &   & X & X & X & V & X & V  & V  & O  & A  \\ \hline
5&Technological infrastructure              &   &   &   &   &   & X & O & V & X & V  & V  & X  & A  \\ \hline
6&Scalability challenges                &   &   &   &   &   &   & O & V & X & V  & V  & X  & A  \\ \hline
7&Cultural and ethical consideration    &   &   &   &   &   &   &   & V & O & V  & O  & O  & A  \\ \hline
8&Acceptance and adoption               &   &   &   &   &   &   &   &   & A & X  & A  & A  & A  \\ \hline
9&Data Integration and management issue &   &   &   &   &   &   &   &   &   & V  & V  & X  & A  \\ \hline
10&Reliability issue                     &   &   &   &   &   &   &   &   &   &    & A  & A  & A  \\ \hline
11&ROI uncertainty                       &   &   &   &   &   &   &   &   &   &    &    & A  & A  \\ \hline
12&Extreme environment issue             &   &   &   &   &   &   &   &   &   &    &    &    & A  \\ \hline
13&Cold chain network                    &   &   &   &   &   &   &   &   &   &    &    &    &    \\ \hline
\end{tabular}
}
\label{tab:table 4.1.1}
\end{table*}

The symbols V, A, X, and O represent the specific sort of connection between the two components $(i, j)$ being examined. The notation ``V" indicates that element $i$ will exert influence on element $j$. ``A" signifies that element $i$ will be impacted by element $j$.The symbol ``X" denotes a bidirectional link, indicating that elements $i$ and $j$ will mutually impact each other. There is no relationship or connection between the elements of the symbol"O".

Table \ref{abbtable} lists IoT adoption barriers using abbreviations for convenience. Each abbreviation provides a concise overview of the main barriers that each one represents. Researchers and stakeholders seeking to overcome and resolve these obstacles in the rapidly evolving field of IoT technology will find the table invaluable.

\begin{table*}[!ht]
\renewcommand\thetable{5}
\centering
\caption{\bf Barrier abbreviation Table}
\resizebox{.55\textwidth}{!}{
\begin{tabular}{|c|c|c|}
\hline
\textbf{Sl no.} & \textbf{Abbreviation} & \textbf{Full form of abbreviation}               \\ \hline
1     & RC           & Regulatory Compliance                   \\ \hline
2     & HIC          & High implementation cost                \\ \hline
3     & LSW          & Lack of skilled workforce               \\ \hline
4     & DSP          & Data security and privacy               \\ \hline
5     & TI           & Technological infrastructure            \\ \hline
6     & SC           & Scalability challenges                  \\ \hline
7     & CEC          & Cultural and ethical   consideration    \\ \hline
8     & AA           & Acceptance and adoption                 \\ \hline
9     & DIM          & Data integration and   management issue \\ \hline
10    & RI           & Reliability issue                       \\ \hline
11    & ROI          & ROI uncertainty                         \\ \hline
12    & EE           & Extreme environment issue               \\ \hline
13    & CCN          & Cold chain network                      \\ \hline
\end{tabular}
}
\label{abbtable}
\end{table*}

\subsubsection{Reachability matrix formation}
The four symbols in \autoref{tab:table 4.1.1} are converted to 0s and 1s to implement SSIM in the first reachability matrix using the rules described in \autoref{ISMstep4}. A square matrix is prepared in Table \ref{tab:table 4.1.2} with binary values that show whether the parts influence one other. The observed barriers are interconnected and have complicated relationships, which can be better understood using the reachability matrix.

\begin{table*}[!ht]
\renewcommand\thetable{6}
\centering
\caption{\bf Initial reachability matrix}
\resizebox{.8\textwidth}{!}{
\begin{tabular}{|lccccccccccccc|}
\hline
\textbf{Barriers}         & \textbf{RC} & \textbf{HIC} & \textbf{LSW} & \textbf{DSP} & \textbf{TI} & \textbf{SC} & \textbf{CEC} & \textbf{AA} & \textbf{DIM} & \textbf{RI} & \textbf{ROI} & \textbf{EE} & \textbf{CCN }\\ \hline
RC        & 1  & 1   & 1   & 1   & 1  & 1  & 1   & 1  & 1   & 1  & 1   & 1  & 1   \\ \hline
HIC       & 0  & 1   & 0   & 0   & 0  & 0  & 0   & 1  & 0   & 1  & 1   & 0  & 0   \\ \hline
LSW       & 0  & 1   & 1   & 0   & 0  & 1  & 1   & 1  & 1   & 1  & 1   & 0  & 0   \\ \hline
DSP       & 0  & 1   & 1   & 1   & 1  & 1  & 1   & 1  & 1   & 1  & 1   & 0  & 0   \\ \hline
TI        & 0  & 1   & 1   & 1   & 1  & 1  & 0   & 1  & 1   & 1  & 1   & 1  & 0   \\ \hline
SC        & 0  & 1   & 0   & 1   & 1  & 1  & 0   & 1  & 1   & 1  & 1   & 1  & 0   \\ \hline
CEC       & 0  & 0   & 1   & 1   & 0  & 0  & 1   & 1  & 0   & 1  & 0   & 0  & 0   \\ \hline
AA        & 0  & 0   & 0   & 0   & 0  & 0  & 0   & 1  & 0   & 1  & 0   & 0  & 0   \\ \hline
DIM       & 0  & 1   & 1   & 1   & 1  & 1  & 0   & 1  & 1   & 1  & 1   & 1  & 0   \\ \hline
RI        & 0  & 0   & 0   & 0   & 0  & 0  & 0   & 1  & 0   & 1  & 0   & 0  & 0   \\ \hline
ROI       & 0  & 1   & 0   & 0   & 0  & 0  & 0   & 1  & 0   & 1  & 1   & 0  & 0   \\ \hline
EE        & 0  & 1   & 0   & 0   & 1  & 1  & 0   & 1  & 1   & 1  & 1   & 1  & 0   \\ \hline
CCN       & 0  & 1   & 1   & 1   & 1  & 1  & 1   & 1  & 1   & 1  & 1   & 1  & 1   \\ \hline
\end{tabular}
}
\label{tab:table 4.1.2}
\end{table*}

\subsubsection{Final reachability matrix formation}
Transitivity was used to create the final reachability matrix, as shown in Table \ref{tab:table 4.1.3}. Following the completion of the analytical phase of the ISM technique, the final reachability matrix, which contains the discovered hierarchical linkages among the system parts, is produced. The final reachability matrix determines the ``driving and dependency power", providing insight into the dynamics of the system. Analyzing ``driving and dependency power," Table \ref{tab:table 4.1.3} classifies parameters into ``dependent, autonomous, linkage, and independent" categories. The barriers that display a significant driving force and dependence are categorized as belonging to the linking group; none of them are classified as autonomous, non-dependent, or dependent.

\begin{table*}[!ht]
\renewcommand\thetable{7}
\caption{\bf Final reachability matrix}
\resizebox{\textwidth}{!}{
\begin{tabular}{|lcccccccccccccc|}
\hline
\textbf{Barriers}         & \textbf{RC} & \textbf{HIC} & \textbf{LSW} & \textbf{DSP} & \textbf{TI} & \textbf{SC} & \textbf{CEC} & \textbf{AA} & \textbf{DIM} & \textbf{RI} & \textbf{ROI} & \textbf{EE} & \textbf{CCN} & \textbf{Driving   Power} \\ \hline
\textbf{RC}               & 1  & 1   & 1   & 1   & 1  & 1  & 1   & 1  & 1   & 1  & 1   & 1  & 1   & 13              \\ \hline
\textbf{HIC}              & 0  & 1   & 0   & 0   & 0  & 0  & 0   & 1  & 0   & 1  & 1   & 0  & 0   & 4               \\ \hline
\textbf{LSW}              & 0  & 1   & 1   & 1*  & 1* & 1  & 1   & 1  & 1   & 1  & 1   & 1* & 0   & 11              \\ \hline
\textbf{DSP}              & 0  & 1   & 1   & 1   & 1  & 1  & 1   & 1  & 1   & 1  & 1   & 1* & 0   & 11              \\ \hline
\textbf{TI }              & 0  & 1   & 1   & 1   & 1  & 1  & 1*  & 1  & 1   & 1  & 1   & 1  & 0   & 11              \\ \hline
\textbf{SC}               & 0  & 1   & 1*  & 1   & 1  & 1  & 1*  & 1  & 1   & 1  & 1   & 1  & 0   & 11              \\ \hline
\textbf{CEC}              & 0  & 1*  & 1   & 1   & 1* & 1* & 1   & 1  & 1*  & 1  & 1*  & 1* & 0   & 11              \\ \hline
\textbf{AA}               & 0  & 0   & 0   & 0   & 0  & 0  & 0   & 1  & 0   & 1  & 0   & 0  & 0   & 2               \\ \hline
\textbf{DIM}              & 0  & 1   & 1   & 1   & 1  & 1  & 1*  & 1  & 1   & 1  & 1   & 1  & 0   & 11              \\ \hline
\textbf{RI}               & 0  & 0   & 0   & 0   & 0  & 0  & 0   & 1  & 0   & 1  & 0   & 0  & 0   & 2               \\ \hline
\textbf{ROI}              & 0  & 1   & 0   & 0   & 0  & 0  & 0   & 1  & 0   & 1  & 1   & 0  & 0   & 4               \\ \hline
\textbf{EE}               & 0  & 1   & 1*  & 1*  & 1  & 1  & 1*  & 1  & 1   & 1  & 1   & 1  & 0   & 11              \\ \hline
\textbf{CCN}              & 0  & 1   & 1   & 1   & 1  & 1  & 1   & 1  & 1   & 1  & 1   & 1  & 1   & 12              \\ \hline
\textbf{Dependence Power} & 1  & 11  & 9   & 9   & 9  & 9  & 9   & 13 & 9   & 13 & 11  & 9  & 2   &                 \\ \hline
\end{tabular}
}
\label{tab:table 4.1.3}
\end{table*}

\subsubsection{Level partition}
Barriers are broken down into three distinct sets: the reachability set, which contains the factor and all other factors it influences; the antecedent set, which contains the factor and all other factors that influence it; and the intersubsection set, which contains all the factors in both sets.

Table \ref{tab:table 4.1.9} provides an overview of the final reachability matrix's detailed reachability set, antecedent set, and inter-subsection set of IoT implementation barriers in CSC. All the barriers were categorized into levels.

\begin{table*}[!ht]
\renewcommand\thetable{8}
\caption{\bf Final level partition}
\resizebox{\textwidth}{!}{
\begin{tabular}{|ccccc|}
\hline
\textbf{Elements $(M_i)$} & \textbf{Reachability Set $R(M_i)$} & \textbf{Antecedent Set $A(N_i)$}                       & \textbf{Intersection Set }  & \textbf{Level }\\ \hline
1            & 1                     & 1                                         & 1                             & 5     \\ \hline
2            & 2, 11                 & 1, 2, 3, 4, 5, 6, 7, 9, 11, 12, 13        & 2, 11                         & 2     \\ \hline
3            & 3, 4, 5, 6, 7, 9, 12  & 1, 3, 4, 5, 6, 7, 9, 12, 13               & 3, 4, 5, 6, 7, 9, 12          & 3     \\ \hline
4            & 3, 4, 5, 6, 7, 9, 12  & 1, 3, 4, 5, 6, 7, 9, 12, 13               & 3, 4, 5, 6, 7, 9, 12          & 3     \\ \hline
5            & 3, 4, 5, 6, 7, 9, 12  & 1, 3, 4, 5, 6, 7, 9, 12, 13               & 3, 4, 5, 6, 7, 9, 12          & 3     \\ \hline
6            & 3, 4, 5, 6, 7, 9, 12  & 1, 3, 4, 5, 6, 7, 9, 12, 13               & 3, 4, 5, 6, 7, 9, 12          & 3     \\ \hline
7            & 3, 4, 5, 6, 7, 9, 12  & 1, 3, 4, 5, 6, 7, 9, 12, 13               & 3, 4, 5, 6, 7, 9, 12          & 3     \\ \hline
8            & 8, 10                 & 1, 2, 3, 4, 5, 6, 7, 8, 9, 10, 11, 12, 13 & 8, 10                         & 1     \\ \hline
9            & 3, 4, 5, 6, 7, 9, 12  & 1, 3, 4, 5, 6, 7, 9, 12, 13               & 3, 4, 5, 6, 7, 9, 12          & 3     \\ \hline
10           & 8, 10                 & 1, 2, 3, 4, 5, 6, 7, 8, 9, 10, 11, 12, 13 & 8, 10                         & 1     \\ \hline
11           & 2, 11                 & 1, 2, 3, 4, 5, 6, 7, 9, 11, 12, 13        & 2, 11                         & 2     \\ \hline
12           & 3, 4, 5, 6, 7, 9, 12  & 1, 3, 4, 5, 6, 7, 9, 12, 13               & 3, 4, 5, 6, 7, 9, 12          & 3     \\ \hline
13           & 13                    & 1, 13                                     & 13                            & 4     \\ \hline
\end{tabular}
}
\label{tab:table 4.1.9}
\end{table*}

\subsubsection{Conical matrix}
The conical matrix in Table \ref{conical_mat} displays the transitivity interactions between adoption barriers. Identifying indirect linkages between components reveals the domino impact of system effects. A more complete picture of the system dynamics and structure is achieved using this matrix, which assists in understanding the wider effect of changing one element on other elements. The conical shape of the final reachability matrix is illustrated in \autoref{fig3}.

\begin{table*}[!ht]
\renewcommand\thetable{9}
\caption{\bf Conical matrix (CM)}
\resizebox{\textwidth}{!}{
\begin{tabular}{|cccccccccccccccc|}
\hline
\textbf{Variables}        & \textbf{8}  & \textbf{10} & \textbf{2}  & \textbf{11} & \textbf{3}  &\textbf{4}  & \textbf{5}  & \textbf{6}  & \textbf{7}  & \textbf{9}  & \textbf{12} & \textbf{13} & \textbf{1} & \textbf{Driving Power} & \textbf{Level} \\ \hline
8                & 1  & 1  & 0  & 0  & 0  & 0  & 0  & 0  & 0  & 0  & 0  & 0  & 0 & 2             & 1     \\ \hline
10               & 1  & 1  & 0  & 0  & 0  & 0  & 0  & 0  & 0  & 0  & 0  & 0  & 0 & 2             & 1     \\ \hline
2                & 1  & 1  & 1  & 1  & 0  & 0  & 0  & 0  & 0  & 0  & 0  & 0  & 0 & 4             & 2     \\ \hline
11               & 1  & 1  & 1  & 1  & 0  & 0  & 0  & 0  & 0  & 0  & 0  & 0  & 0 & 4             & 2     \\ \hline
3                & 1  & 1  & 1  & 1  & 1  & 1* & 1* & 1  & 1  & 1  & 1* & 0  & 0 & 11            & 3     \\ \hline
4                & 1  & 1  & 1  & 1  & 1  & 1  & 1  & 1  & 1  & 1  & 1* & 0  & 0 & 11            & 3     \\ \hline
5                & 1  & 1  & 1  & 1  & 1  & 1  & 1  & 1  & 1* & 1  & 1  & 0  & 0 & 11            & 3     \\ \hline
6                & 1  & 1  & 1  & 1  & 1* & 1  & 1  & 1  & 1* & 1  & 1  & 0  & 0 & 11            & 3     \\ \hline
7                & 1  & 1  & 1* & 1* & 1  & 1  & 1* & 1* & 1  & 1* & 1* & 0  & 0 & 11            & 3     \\ \hline
9                & 1  & 1  & 1  & 1  & 1  & 1  & 1  & 1  & 1* & 1  & 1  & 0  & 0 & 11            & 3     \\ \hline
12               & 1  & 1  & 1  & 1  & 1* & 1* & 1  & 1  & 1* & 1  & 1  & 0  & 0 & 11            & 3     \\ \hline
13               & 1  & 1  & 1  & 1  & 1  & 1  & 1  & 1  & 1  & 1  & 1  & 1  & 0 & 12            & 4     \\ \hline
1                & 1  & 1  & 1  & 1  & 1  & 1  & 1  & 1  & 1  & 1  & 1  & 1  & 1 & 13            & 5     \\ \hline
Dependence Power & 13 & 13 & 11 & 11 & 9  & 9  & 9  & 9  & 9  & 9  & 9  & 2  & 1 &               &       \\ \hline
Level            & 1  & 1  & 2  & 2  & 3  & 3  & 3  & 3  & 3  & 3  & 3  & 4  & 5 &               &       \\ \hline
\end{tabular}
}
\label{conical_mat}
\end{table*}

The hierarchical relationships among the different system parts are better understood using a conical matrix (Table \ref{conical_mat}), which is a tool for creating conical structures. This conical form clarifies the architecture of the system by visually depicting the hierarchical connections.

\begin{figure*}[!ht]
\renewcommand\thefigure{3}
\centering
\includegraphics[width=1\textwidth]{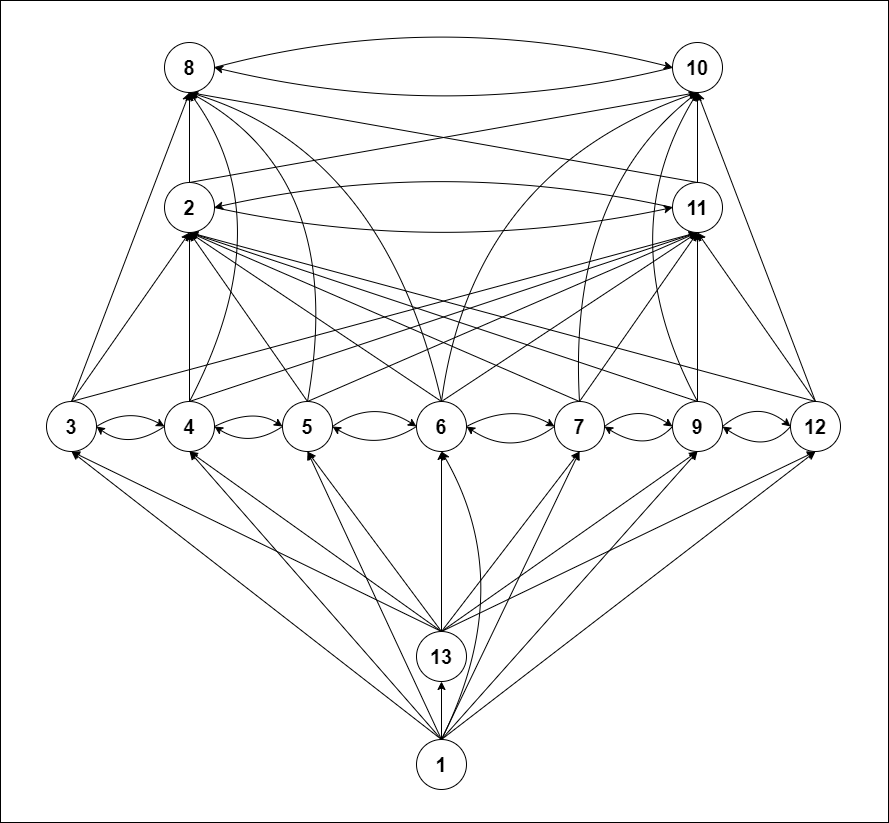}
\caption{\bf Conical form of the final reachability matrix.}
\label{fig3}
\end{figure*}

\subsubsection{MICMAC analysis}
\label{sub:MICMAC}
MICMAC analysis is a useful tool for determining the drive and dependence power of barriers. This analysis was conducted using barriers to divide the structure into separate groups.

\begin{figure*}[!ht]
\renewcommand\thefigure{4}
\centering
\includegraphics[width=1\textwidth]{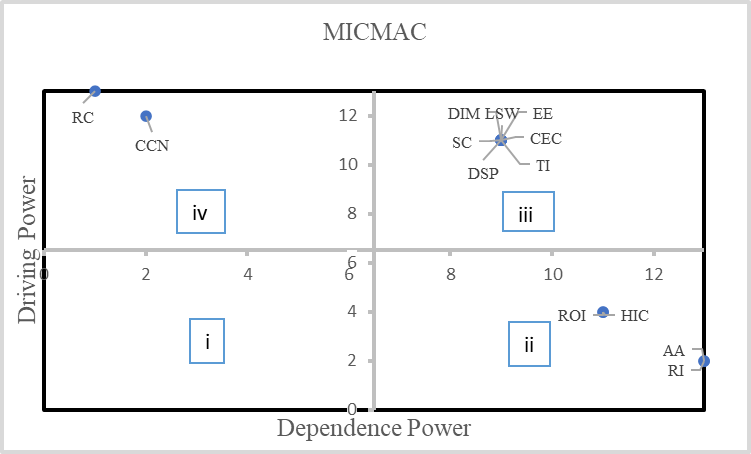}
\caption{\bf MICMAC diagram for barriers.}
\label{fig4}
\end{figure*}

There are four distinct categories of barriers described in Fig \ref{fig4}:\\
\textit{Autonomous barriers:} These barriers were located in the first quadrant. They were not very strong in driving or being dependent.\\
\textit{Dependent barriers:} These barriers are shown in Quadrant-II. The driving power of the subject was weak, whereas the dependence power was strong.\\
\textit{Linkage barriers:} These are inside Quadrant-III and exhibit strong driving and dependent power. In general, the instability of these barriers arises from the interconnectedness of their actions since any action done on one barrier might have consequences for others and could affect the barrier itself.\\
\textit{Independent barriers:} These are shown in Quadrant-IV and exhibit strong driving power but weak dependence power.

\subsubsection{ISM model}
The digraph in Fig \ref{fig5} shows the barriers preventing IoT adoption. It was used to identify the hierarchy levels associated with the implementation of the IoT in cold SC. The conceptual coherence of the produced ISM model was checked, and any required adjustments
were made.

\begin{figure*}[!ht]
\renewcommand\thefigure{5}
\centering
\includegraphics[width=1\textwidth]{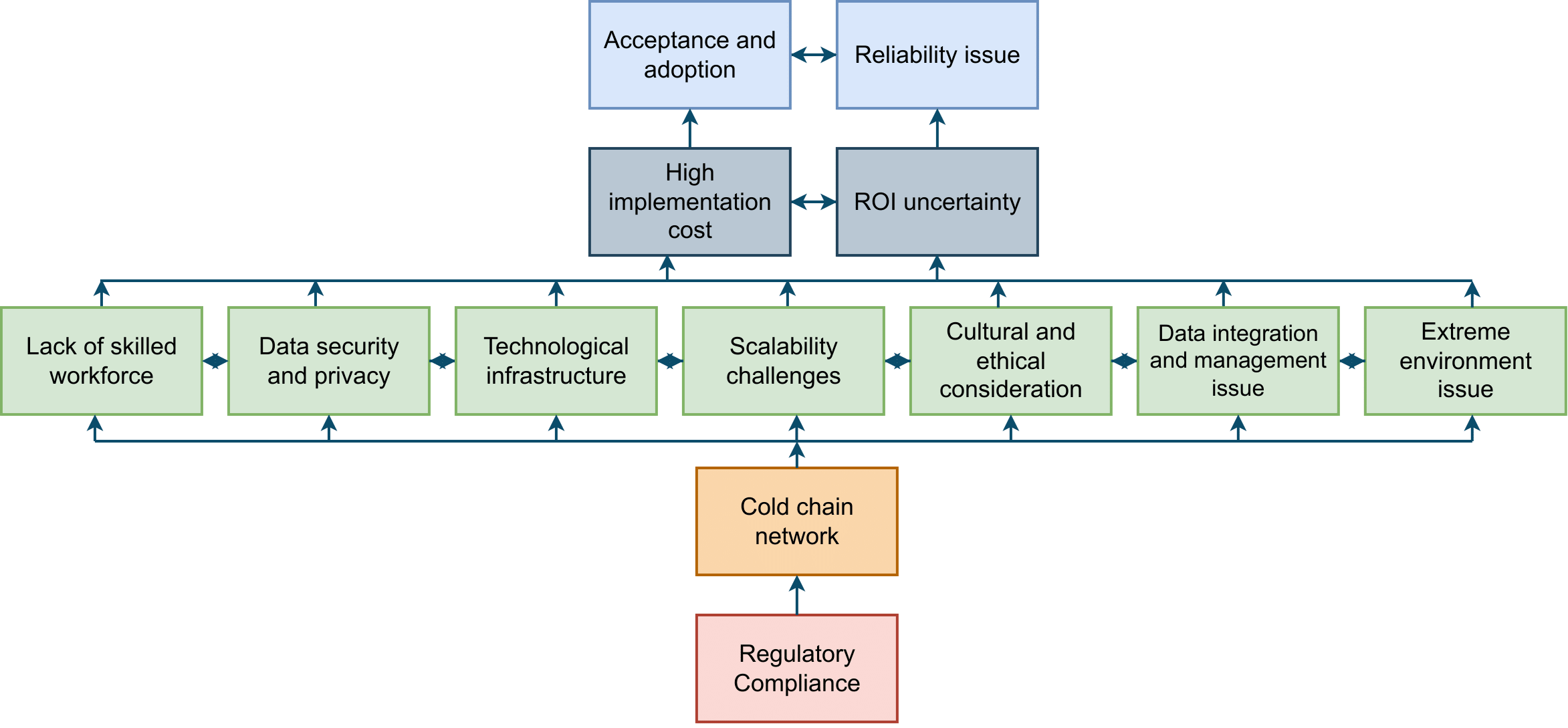}
\caption{\bf Digraph depicting the relationship between barriers.}
\label{fig5}
\end{figure*}

\subsection{DEMATEL application}
\label{sub:dematel}
\subsubsection{Average matrix formation}
Each expert on the panel completed a survey with an integer value between 0 and 4 to indicate the level of impact they noticed between every given pair of obstacles. Using the matrix format of the survey data, the average matrix is determined using \autoref{eqn3.1} in Table \ref{avg_mat}.

\begin{table*}[!ht]
\renewcommand\thetable{10}
\caption{\bf Average matrix}
\resizebox{\textwidth}{!}{
\begin{tabular}{|lccccccccccccc|}
\hline
    & \textbf{RC}    & \textbf{HIC}   & \textbf{LSW}   & \textbf{DSP}   & \textbf{TI}    & \textbf{SC}    & \textbf{CEC}   & \textbf{AA}    & \textbf{DIM}   & \textbf{RI}    & \textbf{ROI}   & \textbf{EE}    & \textbf{CCN}   \\ \hline
\textbf{RC}  & 0.000 & 2.250 & 2.417 & 3.750 & 2.583 & 1.750 & 1.583 & 3.417 & 2.000 & 2.750 & 3.583 & 2.500 & 1.750 \\ \hline
\textbf{HIC} & 0.167 & 0.000 & 0.750 & 0.750 & 0.333 & 0.250 & 0.333 & 2.417 & 0.417 & 3.500 & 3.583 & 0.333 & 1.000 \\ \hline
\textbf{LSW} & 0.167 & 2.333 & 0.000 & 0.167 & 0.833 & 3.500 & 1.750 & 2.833 & 3.500 & 3.500 & 1.833 & 1.083 & 1.167 \\ \hline
\textbf{DSP} & 0.167 & 3.500 & 1.917 & 0.000 & 3.000 & 1.833 & 3.583 & 2.083 & 2.000 & 2.583 & 0.583 & 0.167 & 1.750 \\ \hline
\textbf{TI}  & 0.167 & 2.000 & 1.250 & 2.833 & 0.000 & 1.833 & 0.417 & 1.750 & 3.667 & 3.000 & 2.000 & 3.083 & 0.917 \\ \hline
\textbf{SC}  & 0.167 & 2.917 & 1.083 & 1.333 & 1.833 & 0.000 & 1.000 & 3.083 & 3.167 & 3.250 & 2.000 & 1.667 & 0.667 \\ \hline
\textbf{CEC} & 1.000 & 0.417 & 2.917 & 3.333 & 0.250 & 1.000 & 0.000 & 3.750 & 0.083 & 2.833 & 0.917 & 0.250 & 0.250 \\ \hline
\textbf{AA}  & 0.333 & 0.750 & 1.167 & 1.250 & 0.417 & 0.417 & 2.250 & 0.000 & 0.083 & 2.000 & 1.167 & 0.167 & 0.083 \\ \hline
\textbf{DIM} & 0.250 & 3.000 & 2.333 & 3.167 & 3.583 & 3.250 & 0.000 & 1.750 & 0.000 & 2.833 & 2.083 & 1.000 & 1.500 \\ \hline
\textbf{RI } & 0.000 & 0.333 & 0.333 & 0.250 & 0.333 & 0.500 & 1.000 & 2.167 & 0.250 & 0.000 & 0.333 & 1.083 & 0.250 \\ \hline
\textbf{ROI} & 0.000 & 3.333 & 1.250 & 0.083 & 0.667 & 1.167 & 1.000 & 3.000 & 0.167 & 2.167 & 0.000 & 2.000 & 0.750 \\ \hline
\textbf{EE}  & 0.000 & 3.083 & 0.667 & 0.250 & 2.917 & 2.000 & 0.083 & 2.500 & 0.750 & 1.917 & 2.167 & 0.000 & 0.333 \\ \hline
\textbf{CCN} & 0.083 & 3.250 & 1.917 & 2.083 & 3.083 & 3.000 & 2.583 & 1.750 & 2.833 & 2.000 & 2.917 & 2.917 & 0.000 \\ \hline
\end{tabular}
}
\label{avg_mat}
\end{table*}

\subsubsection{Normalized direct relation matrix formation}
Using \autoref{eqn3.2}, the normalized direct relation matrix is formulated from the average matrix in Table \ref{normalized_mat}. The maximum summation was 32.333, which functions as a divisor, and every cell value from the average matrix was divided by this value.

\begin{table*}[!ht]
\renewcommand\thetable{11}
\caption{\bf Normalised direct relation matrix (D)}
\resizebox{\textwidth}{!}{
\begin{tabular}{|lccccccccccccc|}
\hline
    & \textbf{RC}    & \textbf{HIC}   & \textbf{LSW}   & \textbf{DSP}   & \textbf{TI}    & \textbf{SC}    & \textbf{CEC}   & \textbf{AA}    & \textbf{DIM}   & \textbf{RI}    & \textbf{ROI}   & \textbf{EE}    & \textbf{CCN}   \\ \hline
\textbf{RC}  & 0.000 & 0.070 & 0.075 & 0.116 & 0.080 & 0.054 & 0.049 & 0.106 & 0.062 & 0.085 & 0.111 & 0.077 & 0.054 \\ \hline
\textbf{HIC} & 0.005 & 0.000 & 0.023 & 0.023 & 0.010 & 0.008 & 0.010 & 0.075 & 0.013 & 0.108 & 0.111 & 0.010 & 0.031 \\ \hline
\textbf{LSW} & 0.005 & 0.072 & 0.000 & 0.005 & 0.026 & 0.108 & 0.054 & 0.088 & 0.108 & 0.108 & 0.057 & 0.034 & 0.036 \\ \hline
\textbf{DSP} & 0.005 & 0.108 & 0.059 & 0.000 & 0.093 & 0.057 & 0.111 & 0.064 & 0.062 & 0.080 & 0.018 & 0.005 & 0.054 \\ \hline
\textbf{TI } & 0.005 & 0.062 & 0.039 & 0.088 & 0.000 & 0.057 & 0.013 & 0.054 & 0.113 & 0.093 & 0.062 & 0.095 & 0.028 \\ \hline
\textbf{SC}  & 0.005 & 0.090 & 0.034 & 0.041 & 0.057 & 0.000 & 0.031 & 0.095 & 0.098 & 0.101 & 0.062 & 0.052 & 0.021 \\ \hline
\textbf{CEC} & 0.031 & 0.013 & 0.090 & 0.103 & 0.008 & 0.031 & 0.000 & 0.116 & 0.003 & 0.088 & 0.028 & 0.008 & 0.008 \\ \hline
\textbf{AA}  & 0.010 & 0.023 & 0.036 & 0.039 & 0.013 & 0.013 & 0.070 & 0.000 & 0.003 & 0.062 & 0.036 & 0.005 & 0.003 \\ \hline
\textbf{DIM} & 0.008 & 0.093 & 0.072 & 0.098 & 0.111 & 0.101 & 0.000 & 0.054 & 0.000 & 0.088 & 0.064 & 0.031 & 0.046 \\ \hline
\textbf{RI}  & 0.000 & 0.010 & 0.010 & 0.008 & 0.010 & 0.015 & 0.031 & 0.067 & 0.008 & 0.000 & 0.010 & 0.034 & 0.008 \\ \hline
\textbf{ROI} & 0.000 & 0.103 & 0.039 & 0.003 & 0.021 & 0.036 & 0.031 & 0.093 & 0.005 & 0.067 & 0.000 & 0.062 & 0.023 \\ \hline
\textbf{EE}  & 0.000 & 0.095 & 0.021 & 0.008 & 0.090 & 0.062 & 0.003 & 0.077 & 0.023 & 0.059 & 0.067 & 0.000 & 0.010 \\ \hline
\textbf{CCN} & 0.003 & 0.101 & 0.059 & 0.064 & 0.095 & 0.093 & 0.080 & 0.054 & 0.088 & 0.062 & 0.090 & 0.090 & 0.000 \\ \hline
\end{tabular}
}
\label{normalized_mat}
\end{table*}

\subsubsection{Total relation matrix formation}
The total relation matrix is formulated using \autoref{eqn3.3} in Table \ref{total_mat}. The values in the matrix indicate the degree of relation between the pairs of factors. A higher value suggests a stronger relationship.

\begin{table*}[!ht]
\renewcommand\thetable{12}
\caption{\bf Total relation matrix}
\resizebox{\textwidth}{!}{
\begin{tabular}{|lcccccccccccccc|}
\hline
    & \textbf{RC}    & \textbf{HIC}   & \textbf{LSW}   & \textbf{DSP}   & \textbf{TI}    & \textbf{SC}    & \textbf{CEC}   & \textbf{AA}    & \textbf{DIM}   & \textbf{RI}    & \textbf{ROI}   & \textbf{EE}    & \textbf{CCN}   & \textbf{$R_i$}    \\ \hline
\textbf{RC}  & 0.012 & 0.193 & 0.152 & 0.189 & 0.162 & 0.143 & 0.126 & 0.243 & 0.142 & 0.235 & 0.207 & 0.145 & 0.100 & 2.050 \\ \hline
\textbf{HIC} & 0.009 & 0.048 & 0.053 & 0.049 & 0.039 & 0.041 & 0.044 & 0.130 & 0.039 & 0.162 & 0.143 & 0.040 & 0.047 & 0.843 \\ \hline
\textbf{LSW} & 0.014 & 0.155 & 0.058 & 0.065 & 0.085 & 0.166 & 0.102 & 0.187 & 0.158 & 0.214 & 0.129 & 0.083 & 0.067 & 1.485 \\ \hline
\textbf{DSP} & 0.016 & 0.190 & 0.121 & 0.068 & 0.147 & 0.122 & 0.162 & 0.172 & 0.123 & 0.197 & 0.098 & 0.060 & 0.087 & 1.565 \\ \hline
\textbf{TI}  & 0.013 & 0.157 & 0.097 & 0.141 & 0.070 & 0.125 & 0.068 & 0.158 & 0.168 & 0.204 & 0.137 & 0.143 & 0.065 & 1.546 \\ \hline
\textbf{SC}  & 0.014 & 0.169 & 0.087 & 0.095 & 0.112 & 0.061 & 0.080 & 0.188 & 0.145 & 0.203 & 0.131 & 0.097 & 0.053 & 1.436 \\ \hline
\textbf{CEC} & 0.037 & 0.077 & 0.130 & 0.139 & 0.051 & 0.078 & 0.052 & 0.188 & 0.048 & 0.167 & 0.078 & 0.042 & 0.034 & 1.121 \\ \hline
\textbf{AA}  & 0.015 & 0.058 & 0.062 & 0.063 & 0.035 & 0.040 & 0.093 & 0.050 & 0.026 & 0.109 & 0.064 & 0.026 & 0.017 & 0.660 \\ \hline
\textbf{DIM} & 0.017 & 0.193 & 0.132 & 0.156 & 0.175 & 0.170 & 0.065 & 0.169 & 0.077 & 0.214 & 0.148 & 0.093 & 0.085 & 1.694 \\ \hline
\textbf{RI}  & 0.003 & 0.034 & 0.028 & 0.025 & 0.026 & 0.033 & 0.047 & 0.095 & 0.022 & 0.032 & 0.031 & 0.045 & 0.016 & 0.438 \\ \hline
\textbf{ROI} & 0.006 & 0.149 & 0.071 & 0.035 & 0.053 & 0.071 & 0.063 & 0.155 & 0.038 & 0.136 & 0.050 & 0.089 & 0.041 & 0.959 \\ \hline
\textbf{EE}  & 0.006 & 0.153 & 0.058 & 0.047 & 0.124 & 0.101 & 0.039 & 0.147 & 0.065 & 0.139 & 0.120 & 0.039 & 0.034 & 1.070 \\ \hline
\textbf{CCN} & 0.015 & 0.214 & 0.132 & 0.138 & 0.170 & 0.173 & 0.142 & 0.188 & 0.161 & 0.207 & 0.184 & 0.153 & 0.045 & 1.922 \\ \hline
\textbf{$C_i$}  & 0.178 & 1.790 & 1.180 & 1.212 & 1.249 & 1.325 & 1.082 & 2.071 & 1.211 & 2.219 & 1.522 & 1.055 & 0.691 &       \\ \hline
\end{tabular}
}
\label{total_mat}
\end{table*}

\subsubsection{Identification of cause and effect}

Table \ref{degree} provides information on the degree of influence of various factors related to IoT adoption on cold SC barriers. Each row represents a specific factor, and each column provides different metrics or totals. The row total and column total were calculated using \autoref{eqn3.4} and \autoref{eqn3.5}. Causes have $R_i-C_i$ values greater than 0, whereas the effects have smaller $R_i-C_i$ values.

\begin{table*}[!ht]
\renewcommand\thetable{13}
\caption{\bf Degree of influences}
\resizebox{\textwidth}{!}{
\begin{tabular}{|llll|c|c|c|c|c|}
\hline
\multicolumn{4}{|l|}{\textbf{IoT implementation in cold SC barriers}}  & \textbf{$(R_i)$ Row Total} & \textbf{$(C_i)$ Column Total} & \textbf{$R_i+C_i$}   & \textbf{$R_i-C_i$}    & \textbf{Identify} \\ \hline
\multicolumn{4}{|l|}{Regulatory   Compliance}                           & 2.050          & 0.178             & 2.228  & 1.872  & Cause    \\ \hline
\multicolumn{4}{|l|}{High   implementation cost}                        & 0.843          & 1.790             & 2.633  & -0.947 & Effect   \\ \hline
\multicolumn{4}{|l|}{Lack of skilled   workforce}                       & 1.485          & 1.180             & 2.665  & 0.304  & Cause    \\ \hline
\multicolumn{4}{|l|}{Data security and   privacy}                       & 1.565          & 1.212             & 2.776  & 0.353  & Cause    \\ \hline
\multicolumn{4}{|l|}{Technological   infrastructure}                    & 1.546          & 1.249             & 2.795  & 0.297  & Cause    \\ \hline
\multicolumn{4}{|l|}{Scalability}                                       & 1.436          & 1.325             & 2.761  & 0.110  & Cause    \\ \hline
\multicolumn{4}{|l|}{Cultural and   ethical consideration}              & 1.121          & 1.082             & 2.203  & 0.039  & Cause    \\ \hline
\multicolumn{4}{|l|}{Acceptance and   adoption}                         & 0.660          & 2.071             & 2.731  & -1.412 & Effect   \\ \hline
\multicolumn{4}{|l|}{Data integration   and management issue}           & 1.694          & 1.211             & 2.905  & 0.482  & Cause    \\ \hline
\multicolumn{4}{|l|}{Reliability   issue}                              & 0.438          & 2.219             & 2.657  & -1.781 & Effect   \\ \hline
\multicolumn{4}{|l|}{ROI uncertainty}                                  & 0.959          & 1.522             & 2.481  & -0.563 & Effect   \\ \hline
\multicolumn{4}{|l|}{Extreme   environment issue}                      & 1.070          & 1.055             & 2.125  & 0.015  & Cause    \\ \hline
\multicolumn{4}{|l|}{Cold chain   network}                             & 1.922          & 0.691             & 2.613  & 1.231  & Cause    \\ \hline
\end{tabular}
}
\label{degree}
\end{table*}

\subsubsection{Developing interrelationship map}
A digraph illustrating the influential relationship between the barriers is shown in Fig \ref{fig6}, where the threshold value calculated using \autoref{eqn3.6} is 0.099. Once the threshold value is determined, an interrelationship map is produced. In this map, the values above the threshold reflect the influence of barrier $i$ on barrier $j$. By contrast, when the value is lower than the threshold value, barrier $i$ is influenced by barrier $j$. The arrows demonstrate the influence, demonstrating an important link among the barriers. Values positioned above the neutral horizontal line are categorized as belonging to the cause group, while values positioned below the neutral horizontal line are categorized as belonging to the effect group.

\begin{figure*}[!ht]
\renewcommand\thefigure{6}
\centering
\includegraphics[width=1\textwidth]{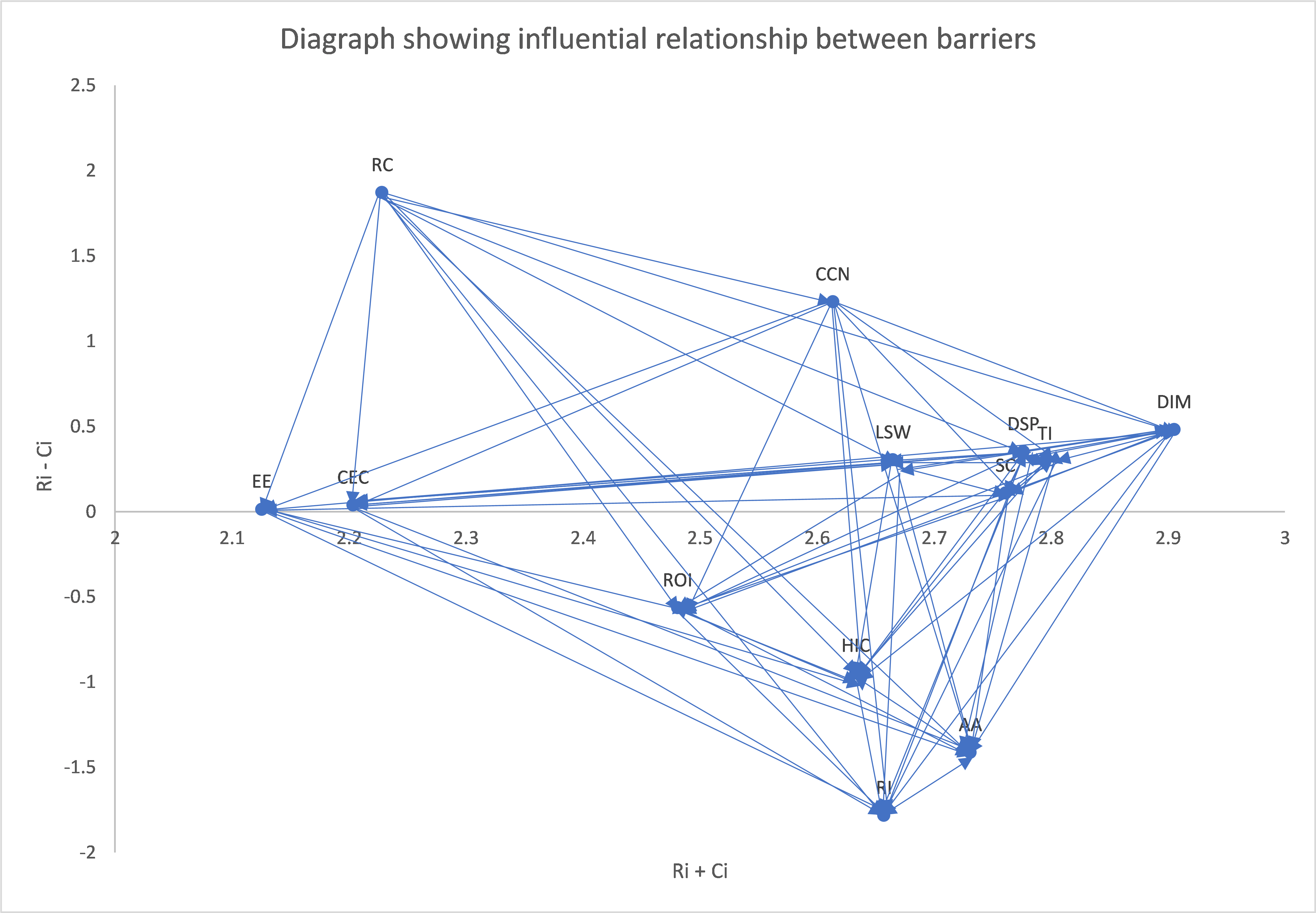}
\caption{\bf Digraph showing the influential relationship between the barriers.}
\label{fig6}
\end{figure*}

\subsubsection{Developing final ranking of attributes}

DIM $>$ TI $>$ DSP $>$ SC $>$ AA $>$ LSW $>$ RI $>$ HIC $>$ CCN $>$ ROI $>$ RC $>$ CEC $>$ EE

Where
Full form of these abbreviations described in \autoref{abbtable}.


\section{Discussions}
\label{sec5}

Numerous interconnected barriers to extensive IoT implementation in cold SC were analyzed in this study. Data from the survey were double-checked to ensure accuracy. The Cronbach's alpha test was used to ensure that the survey results were acceptable and valid. A reliable measurement is ensured by this coefficient, which evaluates the internal consistency of the survey questions. After it was confirmed that the alpha value was above the acceptable range, the subsequent stage of the process was initiated.

To ascertain this, The ISM-MICMAC-DEMATEL method is constructed to determine important barriers, which are divided into three subsections: subsection \ref{sub:ism} for ISM, subsection \ref{sub:MICMAC} for MICMAC and the other subsection \ref{sub:dematel} for DEMATEL.

The ISM model was created to determine the hierarchy levels of IoTBs so that CSC practitioners could see their dependence links and focus on the largest adoption barriers to IoT deployment. This study identified five hierarchical layers to explain the linkages between the chosen IoTBs. The implementation barrier at higher levels exhibits greater susceptibility to the effect of other factors, while components situated at lower levels strongly influence other factors.

Variables classified as high levels, namely Levels 1 and 2, are prone to being influenced by other variables. We identified acceptance, adoption, and reliability issues at the top of the hierarchy, driven by the four hierarchies below. At the second level, we identified two barriers: high implementation cost and ROI uncertainty caused by the three hierarchical layers below it. In the third tier, six interconnected barriers have mutual effects. These barriers include a lack of skilled workforce, data security and privacy, technological infrastructure, scalability challenges, cultural and ethical considerations, data integration and management, and extreme environments. Conversely, components categorized as low levels, namely Levels 4 and 5, can influence the other factors more. The IoT adoption barriers in Level 4 are cold chain networks and Level 5 is regulatory compliance. These factors serve as foundational elements in a broader network of relationships.

The analysis of barriers identifies regulatory compliance as an important factor that hinders the implementation of the IoT. The implementation of government regulations that promote the adoption of IoT will result in a significant transformation of the supply chain system. Infrastructure, environmental factors, a trained workforce, cultural-ethical considerations, scalability, and other factors influence the cold chain network.
Most of the supply chain stages are apprehensive about implementing IoT, as is typical with every new technology. An efficient technological infrastructure is necessary for the successful deployment of the IoT across supply chains, which involves a vast network of linked devices and systems. Adoption of the IoT depends on this infrastructure, which allows integration, data transfer, and communication among components in the supply chain ecosystem. Industries may face challenges with scalability and performance if the infrastructure is insufficient for the IoT. To oversee such a vast and complex system, experts with the necessary level of expertise are required. 
Reducing the challenges associated with IoT adoption requires secure network cloud data. Preventing spyware and attacks is crucial since the number of connected devices and sensitive data is growing. As the nature of threats evolves, more sophisticated security measures and solutions designed for IoT systems may be required.
A massive investment is necessary to hire experts, educate current employees, and build infrastructure. Industries are hesitant to implement IoT due to concerns about the potential lack of return on investment. In order to feel certain about the ROI, small companies need to be presented with a handful of positive examples.

According to the MICMAC analysis, independent barriers are cold chain networks and regulatory compliance; dependent barriers are ROI uncertainty and high implementation costs. The other seven were linkage barriers, and no autonomous barriers existed. Among the many types of barriers, linkage and independent barriers exhibit significant driving forces, making them crucial barriers.

The interdependence of the barriers is depicted in the interrelationship map. Regulatory compliance, lack of skilled workforce, data security and privacy, technological infrastructure, scalability challenges, cultural and ethical considerations, data integration and management issues, extreme environment issues, and cold chain networks are in the cause group. High implementation cost, acceptance and adoption, reliability issues, and ROI uncertainty are in the effect group. Regulatory compliance and cold chain networks were identified as the most important causes, with respective values of 1.8719 and 1.2308. These two barriers drive more in a larger context; thus, improving performance may promote system development. Reliability issues, acceptance and adoption, and high implementation costs are the most important effects influenced by other barriers. These effects were quantified with values of -1.7808, -1.4118, and -94707, respectively. According to the final ranking of attributes, the most important criteria are technological infrastructure, data security and privacy, and data integration and management. The criterion for extreme environments is of the least importance.

To overcome the barriers associated with data integration and management, it is essential to implement standardized software that enables communication between devices on the same system. Priority should be given to privacy and security during IoT implementation, and it is difficult to develop and investigate a gateway that connects devices using multiple protocols. Faculty and industry must collaborate to resolve this challenge. Significant investment and research are needed for solving security and data integration challenges at the ground level.

\subsection{Managerial and practical implications}
Given that the implementation of IoT is in its early stages, the results presented in this paper offer valuable guidance for practitioners. This study has revealed several implications for both managerial and practical considerations, providing insights that can inform decision-making in this evolving field.

\begin{itemize}
    \item This article guides innovative supply chain methods that use new technologies such as IoT. This study provides crucial barriers to understanding supply chain strategies by considering IoT while creating a coordinating system. Managers can also learn about the most prevalent IoT technological difficulties. With additional insight, managers can quickly improve processes, fix issues, and make decisions more informedly.
    \item This research can help CSC managers plan IoT programs to protect perishable goods from packaging and processing to storage, transportation, and distribution by providing precise temperature and humidity data with a data integration system. IoT sensors may warn of temperature changes or equipment breakdown. Managers schedule proactive maintenance to prevent breakdowns and product spoilage. High-quality goods and adequate regulations may benefit from this research.
    \item This study promotes new efforts to provide supply chain stakeholders with a single platform to exchange information to establish a strong system. This study's focus on efficient information transmission made cold SC more trustworthy. IoT devices collect sufficient data to enable managers to observe patterns and trends. Analytics can optimize processes, routes, and resource use.
    \item This study suggests that IoT technology might enhance cold SC product quality, customer demand, and services. If this study is used well, management may replace conventional supply chain linkages with real-time data interchange via radio frequency identification tags, readers, and other embedded electronic devices.
    \item The findings of this study can convince managerial staff to invest in GPS, RFID tags, readers, and other monitoring technologies. Managers will hire suitable candidates to maximize the use of IoT technology using this knowledge. The conclusions of this study may also motivate higher management to expand staff training and development.
\end{itemize}

\subsection{Unique theoretical contribution}

To guarantee a long-term, IoT-based CSC system, this study examined the challenges associated with using the IoT in CSC. Thirteen obstacles were identified with the assistance of stakeholders. According to the results of this study, various stakeholders in CSC require distinct responses. As the transport management system (TMS) is the most influential party, it follows that TMS should take the initiative and rely on policies to guide its actions. These stakeholders initiate the HIC, TI, SC, DIM, EE, and CCN barriers.

Additionally, any and all obstacles may serve as a roadmap for different groups of people engaged in logistics, product processing, quality inspection, and secure packaging. To sum up the theoretical contribution, this study found two kinds of barriers—cause and effect—based on stakeholder theory that prevents the implementation of IoT in CSC systems. The specific barriers to the effective deployment of IoT in CSC may be better understood by exploring each of these categories. Combining insights from management, engineering, logistics, and information systems into a comprehensive plan. This approach facilitates the identification of problems as a whole and the development of comprehensive solutions. This study also used a socio-technical systems approach to examine social-technical interactions inside the CSC. Human traits, organizational culture, and socioeconomic context influence the process of identifying and removing barriers. This model demonstrated how obstacles in one part of a system may affect other parts of the system and how the removal of certain obstacles can cause other benefits to trickle down.

\section{Conclusions and future works}
\label{sec6}
The primary purpose of this study was to identify and examine the barriers that hinder IoT implementation in cold SC. Our research sought to identify and examine the barriers preventing the IoT from being widely used. As a result of IoT's infancy, CSC partners are unaware of their technological requirements and the benefits that accompany them; consequently, they are hesitant to implement it. Significant findings and insights were derived from the ISM-MICMAC-DEMATEL analysis, which clarified the connection between implementation barriers and their effect on CSC. ISM provides hierarchical linkages, but the impact of IoTBs on each other has not been quantified. The DEMATEL method was employed to identify the most crucial IoTBs and overcome the limitations of ISM, which prioritizes the interactions between the selected barriers more than ISM. According to the research results, there are two primary factors that practitioners in CSCs should focus on regulatory compliance and cold chain networks, which serve as a catalyst for advancement by focusing on the barriers hindering development. With MICMAC's element classification based on driving and dependence powers, the interconnections inside a complicated system may be fully understood, and decisions can be made more effectively. According to the DEMATEL attribute ranking, the three most important factors preventing the widespread adoption of the IoT are data integration and management (DIM), technological infrastructure (TI), and data security and privacy (DSP). While these barriers do pose a serious threat to the system, it is vital to remember that they could not be the main causes influencing the dependability and efficiency of IoT deployment as a whole. Regardless of whether they are not the most crucial barriers discovered by DEMATEL analysis, practitioners aiming for long-term viability in IoT deployment should prioritize resolving the most important causes. Challenges with regulatory compliance and cold chain networks are likely to be among these root reasons, as are other essentials for the successful implementation of IoT. The long-term success and stability of IoT deployments may be assured if practitioners concentrate on fixing these underlying issues. On the other hand, where there is a need for rapid or immediate improvement to enhance IoT implementation, it can be practical to tackle the most important obstacles found by DEMATEL first. Although these obstacles may not solve the root problems, they may improve some parts of the IoT deployment, which can lead to rapid success and show stakeholders how beneficial the technology could be.
The efficient implementation of IoT in CSC depends on government rules and a strong cold chain structure. Government engagement may solve data privacy, security, integration, and reliability issues, boosting industry confidence and IoT adoption. Securing and tackling these barriers properly leads to higher reliability in the IoT implementation. The total system becomes more expensive due to the scalability issues and harsh environmental issues. The most important thing to remember is that the long-term advantages will be worth it, even if the initial investment in the IoT is expensive. It will increase the acceptance and adoption of the IoT implementation in CSC.

Practitioners must pay special attention to the highlighted barriers to the adoption of IoT. Hence, this study highlights the present barriers and provides the groundwork for future developments that might revolutionize the use of the IoT in cold SC.
Although our research has shed light on significant insights into IoT implementation barriers within CSCs, it is imperative to acknowledge certain limitations and outline avenues for future exploration. Our primary constraint lies in the indirect nature of our investigation as we could not directly observe industry-specific barriers. Instead, we thoroughly examined existing IoT studies on CSCs. Another limitation of this research is the lack of digital archives, such as Scopus and the ACM Digital Library, in the systematic literature review. However, we selected respectable scientific archives, such as IEEE Xplore, Springer, Elsevier, Taylor \& Francis, Wiley Online Library, ResearchGate, and Science Direct. These repositories provide high-quality research. The references of selected publications were traced using all relevant studies using the backward snowballing method. Identifying thirteen major IoT implementation barriers is the foundation, but it is essential to recognize that additional barriers may surface during real-world implementation. To mitigate potential bias, we sought input from thirty experts. However, the evaluation by these topic experts introduced a degree of subjectivity to our study's final conclusions. Enhancing the robustness of our findings involves soliciting inputs from an extensive array of industry firms. Increased participation would validate our identified barriers and enrich the diversity of perspectives, thereby improving the overall quality of our study. Furthermore, our study's focus on CSC prompts the need for broader investigations into other sectors, such as the food supply chain, agriculture, and perishable goods. Conducting additional research in these areas will provide a more comprehensive contextual understanding and allow for the generalization of our findings beyond the specific case at hand. Additionally, alternative methods should be explored to test the hypotheses regarding the interrelationships among IoT barriers. Although our research has made valuable contributions, addressing these limitations and pursuing future avenues of exploration will undoubtedly refine and extend the scope of our understanding of IoT implementation challenges in various industries.

\section*{Data availability}
All data and related metadata underlying the findings of this study have been deposited in Figshare, an appropriate public data repository. The datasets are accessible at the following DOI: \url{https://doi.org/10.6084/m9.figshare.25826077.v1}.

\nolinenumbers

%
%
%


\end{justify}

\end{document}